\documentclass[aps,pra,reprint,amsmath,amssymb,superscriptaddress,onecolumn,longbibliography, notitlepage,nofootinbib]{revtex4-2}
\usepackage{mathtools}
\usepackage{siunitx}
\usepackage[hidelinks]{hyperref}
\usepackage{svg}
\usepackage[labelfont=bf,justification=justified,format=plain]{caption}
\usepackage{bbm}
\usepackage{graphicx}
\usepackage{floatrow}

\begin{document}
\title{Programmable large-scale simulation of bosonic transport in optical synthetic frequency lattices}

\author{Alen~Senanian}
\email{as3656@cornell.edu}
\affiliation{Department of Physics, Cornell University, Ithaca, NY 14853, USA}
\affiliation{School of Applied and Engineering Physics, Cornell University, Ithaca, NY 14853, USA}
\author{Logan~G.~Wright}
\affiliation{School of Applied and Engineering Physics, Cornell University, Ithaca, NY 14853, USA}
\affiliation{NTT Physics and Informatics Laboratories, NTT Research, Inc., Sunnyvale, CA 94085, USA}
\author{Peter~F.~Wade}
\affiliation{School of Electrical and Computer Engineering, Cornell University, Ithaca, NY 14853, USA}
\author{Hannah~K.~Doyle}
\affiliation{School of Applied and Engineering Physics, Cornell University, Ithaca, NY 14853, USA}
\author{Peter~L.~McMahon}
\email{pmcmahon@cornell.edu}
\affiliation{School of Applied and Engineering Physics, Cornell University, Ithaca, NY 14853, USA}
\affiliation{Kavli Institute at Cornell for Nanoscale Science, Cornell University, Ithaca, NY 14853, USA}

\begin{abstract}
Photonic simulators using synthetic frequency dimensions have enabled flexible experimental analogues of condensed-matter systems, realizing phenomena that are impractical to observe in real-space systems. However, to date such photonic simulators have been limited to small systems suffering from finite-size effects. Here, we present an analog simulator capable of simulating large 2D and 3D lattices, as well as lattices with non-planar connectivity, including a tree lattice that serves as a toy model in quantum gravity. Our demonstration is enabled by the broad bandwidth achievable in photonics, allowing our simulator to realize lattices with over 100,000 lattice sites. We explore these large lattices in a wide range of previously inaccessible regimes by using a novel method to excite arbitrary states. Our work establishes the scalability and flexibility of programmable simulators based on synthetic frequency dimensions in the optical domain. We anticipate that future extensions of this platform will leverage advances in high-bandwidth optoelectronics to support simulations of dynamic, non-equilibrium phases at the scale of millions of lattice sites, and Kerr-frequency-comb technology to simulate models with higher-order interactions, ultimately in regimes and at scales inaccessible to both digital computers and realizable materials. 

\end{abstract}
\maketitle

\section{Introduction}
\label{sec:intro}

Simulations have long been used to understand emergent phenomena in complex many-body systems. Special-purpose analog simulators trade-off the generality of digital implementations for either scalability or access to regimes challenging for digital computers. In this regard, photonic analog simulators~\cite{Peruzzo2010,Lahini2009,Screiber2012,Lin2016,Harris2017,Lin2018,Muniz2019,Wang2020} complement developments in platforms like superconducting circuits~\cite{Hung2021,Karamlou2022} and ultracold atoms~\cite{An2017,Periwal2021} by enabling, in principle, extremely large-scale simulations. Photonic simulation has a long history and has led to the discovery of a variety of phenomena challenging to realize in conventional condensed-matter systems, such as topological phase transitions~\cite{Ozawa2019,Lustig2019,Yang2019,Dutt2020,Lustig2021,Wang2021,Leefmans2022} and non-Hermitian exceptional points~\cite{Regensburger2012,Eichelkraut2013,Hodaei2017,Weidemann2021,Xia2021}, which in turn has led to new photonic devices with applications far beyond basic physical science~\cite{Bandres2018,Solnyshkov2018,Hokmabadi2019,Zeng2020}. 

Although telecommunication technologies routinely utilize the high-bandwidth inherent to optics, harnessing the frequency parallelism of light for large-scale analog simulation has largely remained unexplored. One promising approach is to implement synthetic frequency dimensions~\cite{Bersch2009,Schwartz2013,Ozawa2016,Yuan2016,Bell2017,Qin2018,Dutt2019,Hu2020,Dutt2020,Li2021,Chen2021,Wang2021,YuanDutt2021,Zhao2021,Balcytis2022,Dutt2022}, in which optical frequency modes are mapped to lattice sites to perform bosonic analog simulations. Simulators using synthetic frequency dimensions have been shown to be versatile, implementing synthetic electric and magnetic fields~\cite{Bersch2009,Miyake2013,Yuan2015,Yuan2016,Qin2018,QinYuan2018,Peterson2019,Lee2020,DErrico2021,Li2021,Chen2021,Englebert2021}, non-Hermitian coupling~\cite{WangDutt2021,Wang2021}, and nonlinear interactions~\cite{Englebert2021,Tusnin2020}. So far, however, these have suffered from restricted system sizes, limited programmability, or lacked techniques for probing the underlying dynamics.  

Here, we demonstrate a frequency-mode-based platform that can simulate transport of arbitrary excitations in planar and non-planar optical lattices with up to 100,000 sites -- orders of magnitude greater than achieved previously in programmable simulators~\cite{YuanDutt2021}. By pursuing a dense spectrum with MHz mode spacing, we leverage developments in both optical frequency combs and high-frequency optoelectronics to manipulate and probe a large number of optical frequency modes in a ring cavity. These technologies enable arbitrary frequency encoding of input states, highly programmable lattices, and tools to read out the dynamics of our simulator.  

\begin{figure*}[h]
\centering
    \includegraphics[width=.95\textwidth]{ 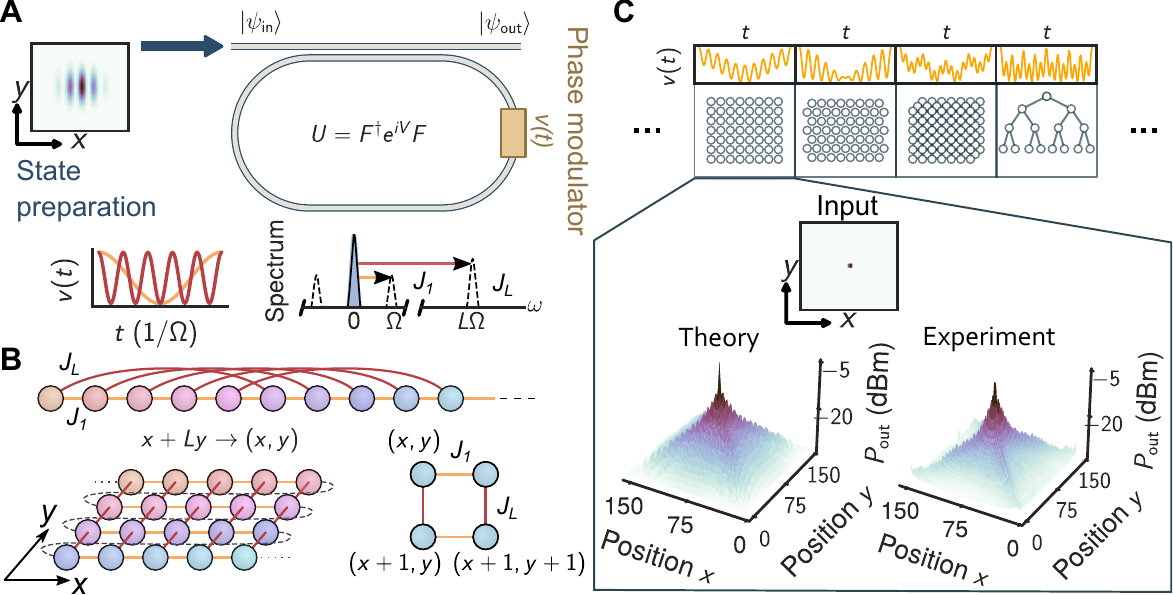}
    \caption{ \textbf{ Simulations of large-scale bosonic transport with programmable photonic simulator.  } (A) Dynamic modulation of a fiber ring resonator couples frequency components of the intracavity field in each roundtrip, represented in the basis of the frequency modes as $U = F^{\dagger} e^{iV} F$. Here, $V \propto \text{diag} (\vec{v})$ is the diagonal voltage operator defining the modulation signal $\vec{v} = (v(t_1),v(t_2),\ldots,v(1/\Omega))$, and $F$ is the discrete Fourier transform. The components of the voltage modulation define the coupling (bottom). By modulating at multiples of the mode spacing ($\Omega$), we only couple long lived modes of the cavity, allowing for injected signals to propagate in frequency for many multiples of the roundtrip time. (B) Engineered long-range coupling maps the one-dimensional spectrum to $L \times L$ two-dimensional lattice with twisted boundary conditions. As $L$ grows large, the lattice approaches a smooth 2D plane. (C) A set of voltage signals defining lattices in 2 and 3 dimensions (top; see main text and Supplementary~Fig.~\ref{fig:modulation_schemes} for details), and the response to a single frequency drive (single-site) for a twisted 2D square lattice with over 20,000 lattice sites compared with tight-binding simulations of a 2D square lattice (bottom). }
    \label{fig:figure1}
\end{figure*}

The class of Hamiltonians that our system is able to simulate is
\begin{equation}
    H = \sum_{i < j} J_{i-j}  a^{\dagger}_i a_j + \text{H.c.}
    \label{eq:hamiltonian}
\end{equation}
Hamiltonians in this class describe non-interacting bosons on lattices with translational invariance. $a^{\dagger}_l$ and $a_l$ are, respectively, the bosonic creation and annihilation operators for the $l$th lattice site. The lattice geometries are defined by the tunneling rates $J_k = \vert J_k \vert e^{i\phi_k}$, which encode the coupling of sites a distance $k$ apart. In the synthetic-frequency-dimensions approach \cite{YuanDutt2021} that we adopt, the lattice operators $a_l$, $a_l^{\dagger}$ are associated with the $l$th frequency mode of an optical cavity; the cavity modes are spaced apart in frequency by $\Omega$ (the free spectral range of the cavity). The tunneling rates between sites $\left( J_k \right)_{k=1,2,3,\ldots}$ are physically realized using a phase modulator within the optical cavity (Fig.~\ref{fig:figure1}A) -- intuitively, the modulator creates optical sidebands at the frequencies contained in the modulation signal $v(t)$ and it is these sidebands that can cause coupling between cavity modes. By setting the frequency components in $v(t)$ appropriately, different lists of tunneling amplitudes $\left( J_k \right)$ can be programmed (Fig.~\ref{fig:figure1}A), which in turn realize different lattice geometries (Fig.~\ref{fig:figure1}C). In our simulator, there is an additional term in the Hamiltonian accounting for the gain/loss balance, but this is kept close to zero (see Supplementary Section~\ref{sec:setup_modeling}).

The main goal of the photonic simulator we present in this work is to be able study the transport of a variety of input excitations in any Hamiltonian in the class defined by Eq.~\eqref{eq:hamiltonian}, where the complex parameters $J_k$ can be programmed arbitrarily -- this allows us to study a diversity of different lattices, including ones that are multidimensional. 

\begin{figure*}[h]
    \centering
    \floatbox[{\capbeside\thisfloatsetup{capbesideposition={right,top},capbesidewidth=8cm}}]{figure}[\FBwidth]
{
\captionsetup{justification=raggedright}
\caption{
\textbf{Optical band-structure measurements of 2D and 3D lattices. } (A) The band structure for a twisted 2D square lattice is measured from the time-domain response of the cavity to scanning single-frequency injection as a function of the detuning $\Delta$~\cite{Dutt2019}, here demonstrated using a toy example with linear lattice size $L = 3$. This time-trace output is sliced up into chunks of length $L$, allowing the reconstruction of a full 2D band structure measured in a single shot (see Supplementary Section~\ref{sec:supp_bs}). (B) Reconstructed band structure for a 2D square lattice with large $L$, comparing analytic (left) with experimental results (right) for $L = 100$. As $L$ grows large, the effect of the twisted boundary condition in the band structure becomes negligible, and the measured band structure approximates that of a regular 2D square lattice. (C) Data of the full bandstructure plotted along slices that connect special points of the Brillouin zone, compared with analytic results for a true 2D square lattice (black). These points, highlighted in the bottom-left, denote locations in momentum space with high symmetry. The density of states $g(E)$ is directly measured by summing the time-domain response (right). Band structures and density of states for (D) 2D triangular lattice ($L = 100$), (E) 3D square lattice ($L = 28$), and (F) 3D hexagonal lattice ($L = 28$).
}\label{fig:figure2}}
{
\includegraphics[width=0.45\textwidth]{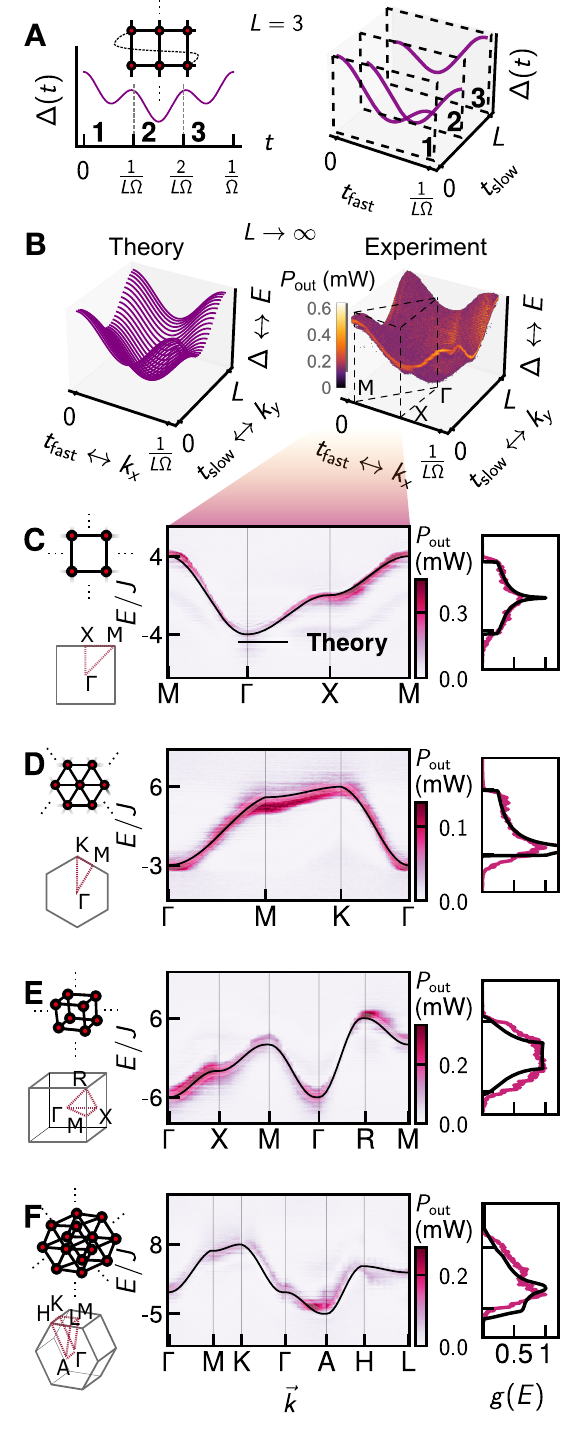}
}
\end{figure*}

The Hamiltonian in Eq.~\eqref{eq:hamiltonian} describes a 1D lattice (with nonlocal couplings) but we can implement effective higher-dimensional lattices by suitably programming the couplings $\left( J_k \right)$ to reflect the local geometry. For example, an effective 2D square lattice can be realized by coupling nearest-neighbors and $L$th nearest-neighbors, i.e., $\left( J_k \right)_{k=1,2,3,\ldots} = \left( J_1,0,\ldots,0,J_L,0,\ldots \right)$  (Fig.~\ref{fig:figure1}B). While this effectively produces a 2D lattice with a twisted boundary condition~\cite{noteforTBCs}, a local excitation with finite lifetime will be insensitive to this twist for a sufficiently large system size. The vanishing effects of the twisted boundary condition can be seen in the steady-state response to a single-site excitation in the comparison shown in Fig.~\ref{fig:figure1}C between our experiment and simulations of a true 2D tight-binding lattice. 

The frequency-multiplexed platform has a convenient encoding of reciprocal space for lattice systems. In the mapping from lattice sites to frequency modes, time maps to momentum~\cite{Dutt2019}. In the same vein, since the Fourier components of the modulation signal define the connectivity, the modulation signal in the time domain defines the band structure. For a 1D lattice, this correspondence is exact: the modulation signal $v(t) = -V_0 \cos(\Omega t)$ couples nearest-neighbor modes, implementing a 1D tight-binding chain with band structure $E(k)  = -J \cos (ka)$. Here the lattice spacing $a$ is identified with the mode spacing $\Omega$, and momentum $k$ with time $t$. More generally, the action of phase modulation on the frequency modes can be expressed as a unitary operator $U = F^{\dagger} e^{i V} F$, where $F$ is the discrete Fourier transform, and $V \propto \text{diag}(\vec{v})$ is a diagonal matrix whose values are proportional to the voltage signal $\vec{v} = (v(t_1),v(t_2),\ldots,v(1/\Omega))$. In our simulator, the operator $U$ implements the time evolution defined by the Hamiltonian in Eq.~\eqref{eq:hamiltonian}. Thus, the modulation signal $v(t)$ defines the time evolution in a diagonal basis, and therefore encodes energy eigenvalues of the lattice, i.e., the band structure. As a consequence, this permits us to encode arbitrary lattices that have with a single-band band structure. Additionally, it provides us direct access to momentum-space lattice measurements \cite{Dutt2019}.  

\begin{figure*}[h]
    \centering
        \includegraphics[width=.9\textwidth]{ 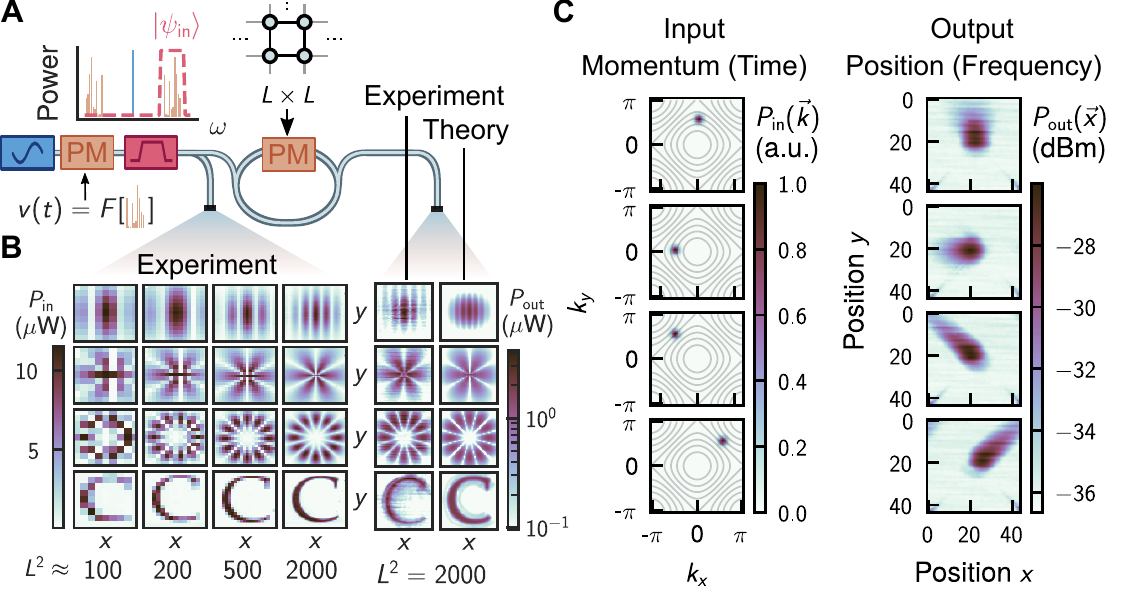}
    \caption{ \textbf{Input state preparation. } (A) Scheme for preparing arbitrary input states: a single frequency tone is modulated with a signal encoding both amplitude and phase of a given state, producing symmetric sidebands (orange spectrum). The initial tone and the unwanted sideband is then rejected with a bandpass filter (red envelope), leaving only the positive sidebands which are sent into the cavity. (B) Experimental measurements of input states for increasing number of modes programmed in the input signal for: a standing wavepacket, an angular wave enveloped with a Gaussian centered at zero, higher angular state enveloped with an offset radial Gaussian, and the Cornell University logo. The steady state outputs of these states for a 2D $L \times L$ square lattice are shown to the right, along with comparision with theory. (C) We excite momentum eigenstates of a 2D square lattice with momenta in various directions enveloped with a Gaussian. Left shows the representation of the input state in momentum space $\vec{k} = (k_x,k_y)$, right shows the experimental steady state in position space $\vec{x} = (x,y)$. Here, local momentum eigenstates are continuously excited at the center, and propagate with a well-defined momentum before decaying. }
    \label{fig:figure3}
\end{figure*}

\begin{figure}[h!]
    \centering
        \includegraphics[width=.95\textwidth]{ 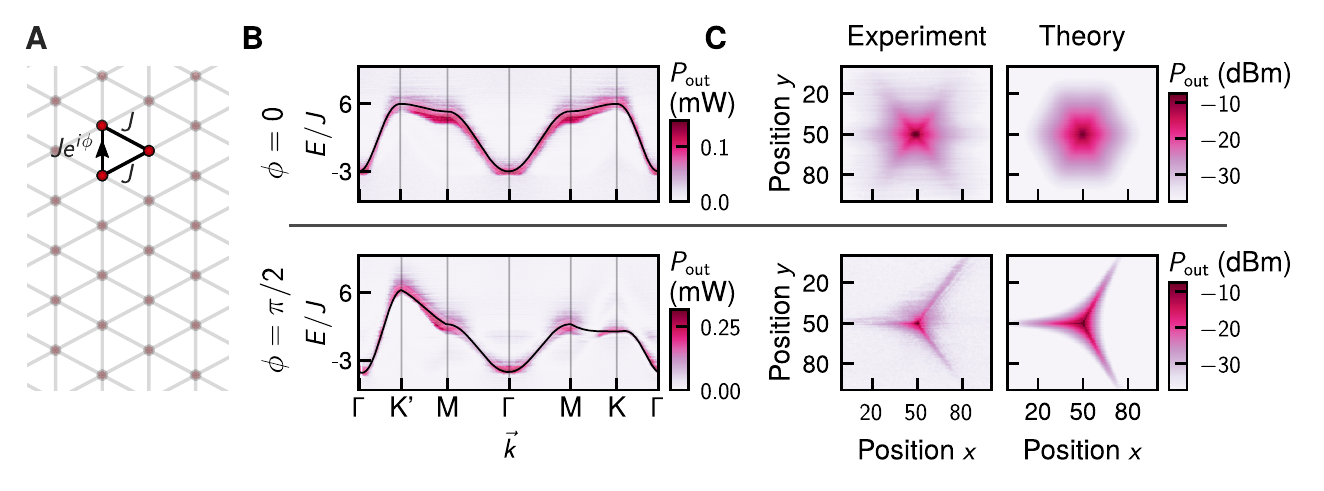}
    \caption{ \textbf{Time-reversal-symmetry breaking in a 2D triangular lattice due to an effective gauge field. } (A) Complex hopping terms induces a nonzero local magnetic flux within a plaquette of a triangular lattice. (B) The introduction of the magnetic field breaks time reversal symmetry, as can be seen in the asymmetry of the K and K’ points in the band structure after performing Pierelis substitution (bottom). (C) Measured steady state spectral response due to a single-site injection under the influence of the synthetic local magnetic field. The presence of the synthetic field leads to a departure from 6-fold symmetry to 3-fold symmetry in the transport. Experimental data (left) is compared with simulations (right).}
    \label{fig:figure4}
\end{figure}

To extend the above momentum-to-time analogy to 2D and 3D, we require the number of modes to grow large enough that the finite-size effects from the twisted boundary condition vanish. For small $L$, slices of the band structure along the slow axis (i.e., the axis corresponding to transport along nearest-neighbors) suffer from an asymmetry near the boundaries of the Brillouin zone (Fig.~\ref{fig:figure2}A). This is due to the twisted boundary conditions, which makes the two directions no longer independent, since $L$ hops along $\Omega$ reaches will reach the same position as a single hop along $L\Omega$. Concretely, the asymmetry in the band structure can be seen by comparing the two-tone signal we use to generate a 2D lattice, $v(t) = -2V_0 \cos (\Omega t) - 2V_0 \cos( L\Omega t)$, and a true 2D tight-binding lattice with nearest-neighbor hopping, which has a band structure $E(\vec{k}) = -2J \cos(k_x a) - 2J \cos(k_y a)$. The latter has two independent reciprocal lattice vectors, $k_x$ and $k_y$. For $L \gg 1$ however, we can rely on a separation of timescales and treat $\Omega' = L\Omega$ as an effective independent degree of freedom. This approach can be readily extended to higher-dimensional lattices, e.g. for a 3D square lattic, $\Omega$, $L\Omega$, and $L^2\Omega$ are the independent degrees of freedom. 

Fig.~\ref{fig:figure2}A outlines how we use the methods introduced in Ref.~\cite{Dutt2019} to measure the band structure of 2D lattice in a single-shot, then slice up the measured band structure in periods of $T_{\text{fast}} = 1/L\Omega$ to reconstruct the 2D full band structure. See Supplementary Section~\ref{sec:supp_bs} for full details on this reconstruction. As $L \rightarrow \infty$, the band structure of our effective 2D square lattice approaches that of a regular 2D square lattice, as seen when comparing Figs.~\ref{fig:figure2}A and B. Slices through high symmetry points of the full band structure are shown in Fig.~\ref{fig:figure2}C-F for a 2D square, 2D triangular, a 3D simple cubic, and a 3D hexagonal lattice, along with the respective density of states for each. Theoretical curves for ordinary tight-binding lattices are shown in black.

High-bandwidth telecommunications optoelectronics enable the study of transport in our platform for arbitrary input states. Our scheme is enabled by 12-GHz electro-optic modulation, summarized schematically in Fig.~\ref{fig:figure3}A. This technique allows us to specify the amplitude and phases of input excitations for up to around 4000 lattice sites, limited primarily by a bandpass filter used in our implementation (see Supplementary Section~\ref{sec:input_state_prep} for details). Figure~\ref{fig:figure3}B shows experimental measurements of various input states, including standing wavepacket eigenstates, angular wavepackets and a Cornell `C'. The right column displays their respective steady-state response. With full control over both amplitude and phase, we are able to excite states with net momentum. Fig.~\ref{fig:figure3}C shows the steady state response of momentum eigenstates of a 2D square lattice enveloped with a Gaussian for a discrete set of nonzero input momenta. The momentum of each input state is labeled by its respective momentum distribution within the Brillouin zone, shown in the left column of Fig.~\ref{fig:figure3}C.

\begin{figure*}[h]
    \centering
        \includegraphics{ 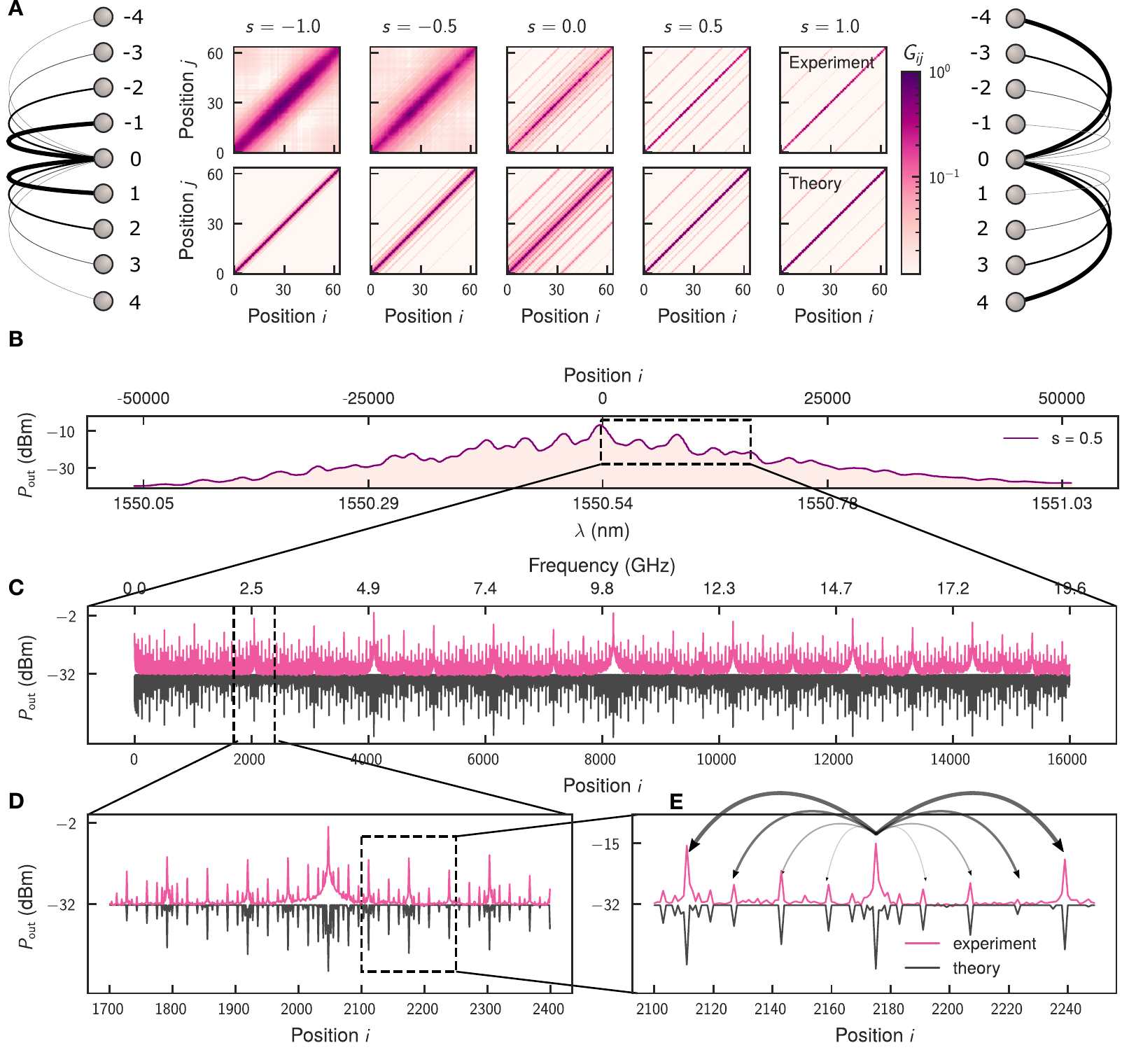}
    \caption{\textbf{Simulations of bosonic transport in a tree-like geometry with a graph comprising over 100,000 sites.} (A) Non-equilibrium correlation measurements for a 1D chain with non-local interactions, characterized by the degree of locality $s$ (see Eq. \eqref{eq:tree}). As $s$ is tuned from $-1$ (left) to $+1$ (right), the correlations transition from locally decaying to treelike \cite{Periwal2021}. Top row is experimental data, bottom row is theory. The lattice cartoons on the left and right schematically show the coupling form for a single lattice site (position 0). (B) Optical spectrum measurement for response to a single site injection with $s = 0.5$, showing measurable steady-state population in $>$ 100,000 lattice sites. (C-E) RF spectrum measurement showing the lattice occupation with single-site resolution of the zoom-in in the full optical spectrum, comparing with simulations for windows of 20 GHz (C), 500 MHz (D), and 100 MHz (E). Theoretical results have been inverted for clarity.}
    \label{fig:figure5}
\end{figure*}

By programming the phases and detunings of the coupling Hamiltonian (Eq.~\eqref{eq:hamiltonian}), we implemented synthetic magnetic and electric fields respectively (see Supplementary Section~\ref{sec:supp_results} for measurements for synthetic electric fields) \cite{Lee2020,DErrico2021,Li2021,Chen2021,Bersch2009,Yuan2016,Qin2018,QinYuan2018,Miyake2013,Yuan2015,Peterson2019}. Figure~\ref{fig:figure4} shows the effect of a synthetic gauge field applied to a 2D triangular lattice, giving rise to a global zero magnetic field, but nonzero local magnetic field. By adding a relative phase along nearest neighbor hoppings, a nonzero magnetic flux going around each plaquette of the triangular lattice is induced. Shown in Fig.~\ref{fig:figure4}B, the addition of this field breaks time-reversal symmetry, which for the triangular and honeycomb lattices, maps the $K$ to $K'$ points \cite{Haldane1988}. This results in a reduction of a 6-fold symmetry to a 3-fold symmetry in the transport of injected light, where propagation of light is prohibited in certain directions, as shown in the heterodyne measurements of the steady-state density in a triangular geometry with 10,000 lattice sites (Fig.~\ref{fig:figure4}). This time-reversal-symmetry breaking with local non-zero fields is one key ingredient in observations of the quantum valley Hall effect seen in honeycomb lattices~\cite{Haldane1988,Mak2014,JimnezGaln2020}.

In addition to lattices found in traditional condensed-matter systems, our photonic simulator is capable of simulating systems not realizable in crystalline materials. Systems with non-planar connectivity are particularly interesting given their realization in solid state systems are impractical, yet contain rich physics. 

~\textcite{Periwal2021} recently experimentally demonstrated a simulation of a graph with exotic long-range connectivity given by
\begin{equation}
    J_{i - j} \propto \begin{cases} 
          \vert i - j \vert ^s & \vert i - j \vert = 2^n, n \in \mathbb{Z}^+\\
          0 & \text{otherwise}. 
       \end{cases}
       \label{eq:tree}
\end{equation}
This describes a system that can be continuously changed, using the parameter $s$, from an Archimedean-geometry regime in which correlations between sites decay with Archimedean distance $\vert i - j \vert $, to a non-Archimedean-geometry regime in which the correlations between sites have a treelike structure. The hierarchical geometry of this treelike system is a toy model for $p$-adic AdS/CFT correspondence~\cite{Bentsen2019} studied
previously using atomic ensembles in an optical lattice with 16 sites~\cite{Periwal2021}. We experimentally show this transition in Fig.~\ref{fig:figure5}A in the measurements of correlations of the lattice as $s$ is tuned (see Supplementary~Section~\ref{sec:supp_results} for details on the correlation measurements and another example with a 1D lattice). Near the transition, at $s = 0.5$, the lattice exhibits both strong local and nonlocal connectivity, resulting in dense yet extremely large lattices, shown in both optical spectrum measurements in Fig.~\ref{fig:figure5}B as well as RF spectrum measurements in Fig.~\ref{fig:figure5}C. 

\section{Discussion}
\label{sec:discussion}

While some graphs, such as the tree-like example depicted in Fig.~\ref{fig:figure5}, result in steady-state occupations that span 100,000 or more lattice sites, quantifying the absolute size of our simulator requires some nuance. On one hand, based on the dispersion and bandwidth of the elements inside the cavity, we believe the lattices we simulate span several THz, corresponding to millions of lattice sites. On the other hand, as in real systems, local excitations in locally connected lattices will rarely explore most of this large lattice before their amplitude decays below the noise floor of our detectors. In these lattices, the steady-state response of a local injection is detectable at most $10^4$ lattice sites in the vicinity of the injected wavepacket.

The examples here cover only a narrow swath of the bosonic physics that can be simulated with frequency-domain coupling of photonic modes. By optimization of intracavity dispersion, loss, and power, observing dynamics on lattices spanning millions or more sites is feasible. By adding intracavity amplitude modulators, non-Hermitian Hamiltonians can be realized~\cite{Wang2021}. By adding multiple spatial modes or coupled cavities, multi-band physics~\cite{Dutt2020}, defects, and hard lattice edges may be simulated~\cite{Dutt2022}, enabling the study of much more complex topological phenomena. By varying modulations over multiple cavity periods, dynamical, non-equilibrium phases~\cite{Battiston2021,Disa2021} and other transient phenomena may be realized. In unmodulated cavities, Kerr nonlinearities give rise to locked combs defined by dissipative cavity solitons~\cite{Kippenberg2018}. Intermodal interactions lead to four-mode interactions~\cite{Tusnin2020,Englebert2021}, which could be systematically programmed via reconfigurable intracavity dispersion, spectral loss, and/or additional mode families~\cite{Wright2017,Karpov2019}. Such platforms would allow new regimes of intricately tailored frequency combs, and would permit photonic frequency-mode simulators to explore the physics of models with higher-order interactions, a regime which remains a challenging frontier of network science~\cite{Battiston2021}. Photonic simulators have been demonstrated to be robust platforms for exploring condensed-matter phenomena, some of which had previously been inaccessible. Frequency-domain photonic simulators benefit from programmability and the potential for large scale, with near-term prospects to scale to sizes that would be impractical to simulate with conventional computers, and in turn may enable the study of emergent behavior that cannot currently be explored in any setting. 

\section*{Data and code availability}
All data generated and code used in this work is
available at: \url{https://doi.org/10.5281/zenodo.6959554}

\section*{Author Contributions}

A.S., L.G.W. and P.L.M. developed the concept. A.S. and L.G.W. built the experimental setup, with early contributions from H.K.D. A.S. performed the experiments, the data analysis, and the numerical simulations (theory). P.F.W. performed experimental data collection. A.S., L.G.W. and P.L.M. wrote the manuscript. L.G.W. and P.L.M. supervised the project.

\section*{Acknowledgements}

P.L.M. gratefully acknowledges financial support from a David and Lucile Packard Foundation Fellowship, and also acknowledges membership of the CIFAR Quantum Information Science Program as an Azrieli Global Scholar. The authors wish to thank NTT Research for their financial and technical support. Portions of this work were supported by the National Science Foundation (award CCF-1918549). We acknowledge helpful discussions with David Hathcock, Erich Mueller, Sridhar Prabhu, Eliott Rosenberg, and members of the NTT PHI Lab / NSF Expeditions research collaboration. We also thank Avik Dutt for helpful discussions and for feedback on a draft of the manuscript. We thank Michael Buttolph for assistance with fiber splicing, and Valene Tjong for contributing instrumentation-control code. 

\bibliography{references}

\begin{thebibliography}{63}%
\makeatletter
\providecommand \@ifxundefined [1]{%
 \@ifx{#1\undefined}
}%
\providecommand \@ifnum [1]{%
 \ifnum #1\expandafter \@firstoftwo
 \else \expandafter \@secondoftwo
 \fi
}%
\providecommand \@ifx [1]{%
 \ifx #1\expandafter \@firstoftwo
 \else \expandafter \@secondoftwo
 \fi
}%
\providecommand \natexlab [1]{#1}%
\providecommand \enquote  [1]{``#1''}%
\providecommand \bibnamefont  [1]{#1}%
\providecommand \bibfnamefont [1]{#1}%
\providecommand \citenamefont [1]{#1}%
\providecommand \href@noop [0]{\@secondoftwo}%
\providecommand \href [0]{\begingroup \@sanitize@url \@href}%
\providecommand \@href[1]{\@@startlink{#1}\@@href}%
\providecommand \@@href[1]{\endgroup#1\@@endlink}%
\providecommand \@sanitize@url [0]{\catcode `\\12\catcode `\$12\catcode
  `\&12\catcode `\#12\catcode `\^12\catcode `\_12\catcode `\%12\relax}%
\providecommand \@@startlink[1]{}%
\providecommand \@@endlink[0]{}%
\providecommand \url  [0]{\begingroup\@sanitize@url \@url }%
\providecommand \@url [1]{\endgroup\@href {#1}{\urlprefix }}%
\providecommand \urlprefix  [0]{URL }%
\providecommand \Eprint [0]{\href }%
\providecommand \doibase [0]{https://doi.org/}%
\providecommand \selectlanguage [0]{\@gobble}%
\providecommand \bibinfo  [0]{\@secondoftwo}%
\providecommand \bibfield  [0]{\@secondoftwo}%
\providecommand \translation [1]{[#1]}%
\providecommand \BibitemOpen [0]{}%
\providecommand \bibitemStop [0]{}%
\providecommand \bibitemNoStop [0]{.\EOS\space}%
\providecommand \EOS [0]{\spacefactor3000\relax}%
\providecommand \BibitemShut  [1]{\csname bibitem#1\endcsname}%
\let\auto@bib@innerbib\@empty
\bibitem [{\citenamefont {Peruzzo}\ \emph {et~al.}(2010)\citenamefont
  {Peruzzo}, \citenamefont {Lobino}, \citenamefont {Matthews}, \citenamefont
  {Matsuda}, \citenamefont {Politi}, \citenamefont {Poulios}, \citenamefont
  {Zhou}, \citenamefont {Lahini}, \citenamefont {Ismail}, \citenamefont
  {W\"{o}rhoff}, \citenamefont {Bromberg}, \citenamefont {Silberberg},
  \citenamefont {Thompson},\ and\ \citenamefont {O'Brien}}]{Peruzzo2010}%
  \BibitemOpen
  \bibfield  {author} {\bibinfo {author} {\bibfnamefont {A.}~\bibnamefont
  {Peruzzo}}, \bibinfo {author} {\bibfnamefont {M.}~\bibnamefont {Lobino}},
  \bibinfo {author} {\bibfnamefont {J.~C.~F.}\ \bibnamefont {Matthews}},
  \bibinfo {author} {\bibfnamefont {N.}~\bibnamefont {Matsuda}}, \bibinfo
  {author} {\bibfnamefont {A.}~\bibnamefont {Politi}}, \bibinfo {author}
  {\bibfnamefont {K.}~\bibnamefont {Poulios}}, \bibinfo {author} {\bibfnamefont
  {X.-Q.}\ \bibnamefont {Zhou}}, \bibinfo {author} {\bibfnamefont
  {Y.}~\bibnamefont {Lahini}}, \bibinfo {author} {\bibfnamefont
  {N.}~\bibnamefont {Ismail}}, \bibinfo {author} {\bibfnamefont
  {K.}~\bibnamefont {W\"{o}rhoff}}, \bibinfo {author} {\bibfnamefont
  {Y.}~\bibnamefont {Bromberg}}, \bibinfo {author} {\bibfnamefont
  {Y.}~\bibnamefont {Silberberg}}, \bibinfo {author} {\bibfnamefont {M.~G.}\
  \bibnamefont {Thompson}},\ and\ \bibinfo {author} {\bibfnamefont {J.~L.}\
  \bibnamefont {O'Brien}},\ }\bibfield  {title} {\bibinfo {title} {Quantum
  walks of correlated photons},\ }\href
  {https://doi.org/10.1126/science.1193515} {\bibfield  {journal} {\bibinfo
  {journal} {Science}\ }\textbf {\bibinfo {volume} {329}} (\bibinfo {year}
  {2010})}\BibitemShut {NoStop}%
\bibitem [{\citenamefont {Lahini}\ \emph {et~al.}(2009)\citenamefont {Lahini},
  \citenamefont {Pugatch}, \citenamefont {Pozzi}, \citenamefont {Sorel},
  \citenamefont {Morandotti}, \citenamefont {Davidson},\ and\ \citenamefont
  {Silberberg}}]{Lahini2009}%
  \BibitemOpen
  \bibfield  {author} {\bibinfo {author} {\bibfnamefont {Y.}~\bibnamefont
  {Lahini}}, \bibinfo {author} {\bibfnamefont {R.}~\bibnamefont {Pugatch}},
  \bibinfo {author} {\bibfnamefont {F.}~\bibnamefont {Pozzi}}, \bibinfo
  {author} {\bibfnamefont {M.}~\bibnamefont {Sorel}}, \bibinfo {author}
  {\bibfnamefont {R.}~\bibnamefont {Morandotti}}, \bibinfo {author}
  {\bibfnamefont {N.}~\bibnamefont {Davidson}},\ and\ \bibinfo {author}
  {\bibfnamefont {Y.}~\bibnamefont {Silberberg}},\ }\bibfield  {title}
  {\bibinfo {title} {Observation of a localization transition in quasiperiodic
  photonic lattices},\ }\href {https://doi.org/10.1103/PhysRevLett.103.013901}
  {\bibfield  {journal} {\bibinfo  {journal} {Physical Review Letters}\
  }\textbf {\bibinfo {volume} {103}},\ \bibinfo {pages} {013901} (\bibinfo
  {year} {2009})}\BibitemShut {NoStop}%
\bibitem [{\citenamefont {Schreiber}\ \emph {et~al.}(2012)\citenamefont
  {Schreiber}, \citenamefont {Gábris}, \citenamefont {Rohde}, \citenamefont
  {Laiho}, \citenamefont {Štefaňák}, \citenamefont {Potoček}, \citenamefont
  {Hamilton}, \citenamefont {Jex},\ and\ \citenamefont
  {Silberhorn}}]{Screiber2012}%
  \BibitemOpen
  \bibfield  {author} {\bibinfo {author} {\bibfnamefont {A.}~\bibnamefont
  {Schreiber}}, \bibinfo {author} {\bibfnamefont {A.}~\bibnamefont {Gábris}},
  \bibinfo {author} {\bibfnamefont {P.~P.}\ \bibnamefont {Rohde}}, \bibinfo
  {author} {\bibfnamefont {K.}~\bibnamefont {Laiho}}, \bibinfo {author}
  {\bibfnamefont {M.}~\bibnamefont {Štefaňák}}, \bibinfo {author}
  {\bibfnamefont {V.}~\bibnamefont {Potoček}}, \bibinfo {author}
  {\bibfnamefont {C.}~\bibnamefont {Hamilton}}, \bibinfo {author}
  {\bibfnamefont {I.}~\bibnamefont {Jex}},\ and\ \bibinfo {author}
  {\bibfnamefont {C.}~\bibnamefont {Silberhorn}},\ }\bibfield  {title}
  {\bibinfo {title} {A 2d quantum walk simulation of two-particle dynamics},\
  }\href {https://doi.org/10.1126/science.1218448} {\bibfield  {journal}
  {\bibinfo  {journal} {Science}\ }\textbf {\bibinfo {volume} {336}} (\bibinfo
  {year} {2012})}\BibitemShut {NoStop}%
\bibitem [{\citenamefont {Lin}\ \emph {et~al.}(2016)\citenamefont {Lin},
  \citenamefont {Xiao}, \citenamefont {Yuan},\ and\ \citenamefont
  {Fan}}]{Lin2016}%
  \BibitemOpen
  \bibfield  {author} {\bibinfo {author} {\bibfnamefont {Q.}~\bibnamefont
  {Lin}}, \bibinfo {author} {\bibfnamefont {M.}~\bibnamefont {Xiao}}, \bibinfo
  {author} {\bibfnamefont {L.}~\bibnamefont {Yuan}},\ and\ \bibinfo {author}
  {\bibfnamefont {S.}~\bibnamefont {Fan}},\ }\bibfield  {title} {\bibinfo
  {title} {Photonic {Weyl} point in a two-dimensional resonator lattice with a
  synthetic frequency dimension},\ }\href {https://doi.org/10.1038/ncomms13731}
  {\bibfield  {journal} {\bibinfo  {journal} {Nature Communications}\ }\textbf
  {\bibinfo {volume} {7}} (\bibinfo {year} {2016})}\BibitemShut {NoStop}%
\bibitem [{\citenamefont {Harris}\ \emph {et~al.}(2017)\citenamefont {Harris},
  \citenamefont {Steinbrecher}, \citenamefont {Prabhu}, \citenamefont {Lahini},
  \citenamefont {Mower}, \citenamefont {Bunandar}, \citenamefont {Chen},
  \citenamefont {Wong}, \citenamefont {Baehr-Jones}, \citenamefont {Hochberg}
  \emph {et~al.}}]{Harris2017}%
  \BibitemOpen
  \bibfield  {author} {\bibinfo {author} {\bibfnamefont {N.~C.}\ \bibnamefont
  {Harris}}, \bibinfo {author} {\bibfnamefont {G.~R.}\ \bibnamefont
  {Steinbrecher}}, \bibinfo {author} {\bibfnamefont {M.}~\bibnamefont
  {Prabhu}}, \bibinfo {author} {\bibfnamefont {Y.}~\bibnamefont {Lahini}},
  \bibinfo {author} {\bibfnamefont {J.}~\bibnamefont {Mower}}, \bibinfo
  {author} {\bibfnamefont {D.}~\bibnamefont {Bunandar}}, \bibinfo {author}
  {\bibfnamefont {C.}~\bibnamefont {Chen}}, \bibinfo {author} {\bibfnamefont
  {F.~N.}\ \bibnamefont {Wong}}, \bibinfo {author} {\bibfnamefont
  {T.}~\bibnamefont {Baehr-Jones}}, \bibinfo {author} {\bibfnamefont
  {M.}~\bibnamefont {Hochberg}}, \emph {et~al.},\ }\bibfield  {title} {\bibinfo
  {title} {Quantum transport simulations in a programmable nanophotonic
  processor},\ }\href {http://dx.doi.org/10.1038/nphoton.2017.95} {\bibfield
  {journal} {\bibinfo  {journal} {Nature Photonics}\ }\textbf {\bibinfo
  {volume} {11}},\ \bibinfo {pages} {447} (\bibinfo {year} {2017})}\BibitemShut
  {NoStop}%
\bibitem [{\citenamefont {Lin}\ \emph {et~al.}(2018)\citenamefont {Lin},
  \citenamefont {Sun}, \citenamefont {Xiao}, \citenamefont {Zhang},\ and\
  \citenamefont {Fan}}]{Lin2018}%
  \BibitemOpen
  \bibfield  {author} {\bibinfo {author} {\bibfnamefont {Q.}~\bibnamefont
  {Lin}}, \bibinfo {author} {\bibfnamefont {X.-Q.}\ \bibnamefont {Sun}},
  \bibinfo {author} {\bibfnamefont {M.}~\bibnamefont {Xiao}}, \bibinfo {author}
  {\bibfnamefont {S.-C.}\ \bibnamefont {Zhang}},\ and\ \bibinfo {author}
  {\bibfnamefont {S.}~\bibnamefont {Fan}},\ }\bibfield  {title} {\bibinfo
  {title} {A three-dimensional photonic topological insulator using a
  two-dimensional ring resonator lattice with a synthetic frequency
  dimension},\ }\href {https://doi.org/10.1126/sciadv.aat2774} {\bibfield
  {journal} {\bibinfo  {journal} {Science Advances}\ }\textbf {\bibinfo
  {volume} {4}} (\bibinfo {year} {2018})}\BibitemShut {NoStop}%
\bibitem [{\citenamefont {Muniz}\ \emph {et~al.}(2019)\citenamefont {Muniz},
  \citenamefont {Wimmer}, \citenamefont {Bisianov}, \citenamefont
  {Morandotti},\ and\ \citenamefont {Peschel}}]{Muniz2019}%
  \BibitemOpen
  \bibfield  {author} {\bibinfo {author} {\bibfnamefont {A.~L.~M.}\
  \bibnamefont {Muniz}}, \bibinfo {author} {\bibfnamefont {M.}~\bibnamefont
  {Wimmer}}, \bibinfo {author} {\bibfnamefont {A.}~\bibnamefont {Bisianov}},
  \bibinfo {author} {\bibfnamefont {R.}~\bibnamefont {Morandotti}},\ and\
  \bibinfo {author} {\bibfnamefont {U.}~\bibnamefont {Peschel}},\ }\bibfield
  {title} {\bibinfo {title} {Collapse on the line – how synthetic dimensions
  influence nonlinear effects},\ }\href
  {https://doi.org/10.1038/s41598-019-46060-8} {\bibfield  {journal} {\bibinfo
  {journal} {Scientific Reports}\ }\textbf {\bibinfo {volume} {9}},\ \bibinfo
  {pages} {9518} (\bibinfo {year} {2019})}\BibitemShut {NoStop}%
\bibitem [{\citenamefont {Wang}\ \emph {et~al.}(2020)\citenamefont {Wang},
  \citenamefont {Zheng}, \citenamefont {Chen}, \citenamefont {Huang},
  \citenamefont {Kartashov}, \citenamefont {Torner}, \citenamefont {Konotop},\
  and\ \citenamefont {Ye}}]{Wang2020}%
  \BibitemOpen
  \bibfield  {author} {\bibinfo {author} {\bibfnamefont {P.}~\bibnamefont
  {Wang}}, \bibinfo {author} {\bibfnamefont {Y.}~\bibnamefont {Zheng}},
  \bibinfo {author} {\bibfnamefont {X.}~\bibnamefont {Chen}}, \bibinfo {author}
  {\bibfnamefont {C.}~\bibnamefont {Huang}}, \bibinfo {author} {\bibfnamefont
  {Y.~V.}\ \bibnamefont {Kartashov}}, \bibinfo {author} {\bibfnamefont
  {L.}~\bibnamefont {Torner}}, \bibinfo {author} {\bibfnamefont {V.~V.}\
  \bibnamefont {Konotop}},\ and\ \bibinfo {author} {\bibfnamefont
  {F.}~\bibnamefont {Ye}},\ }\bibfield  {title} {\bibinfo {title} {Localization
  and delocalization of light in photonic moiré lattices},\ }\href
  {https://doi.org/10.1038/s41586-019-1851-6} {\bibfield  {journal} {\bibinfo
  {journal} {Nature}\ }\textbf {\bibinfo {volume} {577}} (\bibinfo {year}
  {2020})}\BibitemShut {NoStop}%
\bibitem [{\citenamefont {Hung}\ \emph {et~al.}(2021)\citenamefont {Hung},
  \citenamefont {Busnaina}, \citenamefont {Chang}, \citenamefont {Vadiraj},
  \citenamefont {Nsanzineza}, \citenamefont {Solano}, \citenamefont {Alaeian},
  \citenamefont {Rico},\ and\ \citenamefont {Wilson}}]{Hung2021}%
  \BibitemOpen
  \bibfield  {author} {\bibinfo {author} {\bibfnamefont {J.~S.~C.}\
  \bibnamefont {Hung}}, \bibinfo {author} {\bibfnamefont {J.~H.}\ \bibnamefont
  {Busnaina}}, \bibinfo {author} {\bibfnamefont {C.~W.~S.}\ \bibnamefont
  {Chang}}, \bibinfo {author} {\bibfnamefont {A.~M.}\ \bibnamefont {Vadiraj}},
  \bibinfo {author} {\bibfnamefont {I.}~\bibnamefont {Nsanzineza}}, \bibinfo
  {author} {\bibfnamefont {E.}~\bibnamefont {Solano}}, \bibinfo {author}
  {\bibfnamefont {H.}~\bibnamefont {Alaeian}}, \bibinfo {author} {\bibfnamefont
  {E.}~\bibnamefont {Rico}},\ and\ \bibinfo {author} {\bibfnamefont {C.~M.}\
  \bibnamefont {Wilson}},\ }\bibfield  {title} {\bibinfo {title} {Quantum
  simulation of the bosonic {C}reutz ladder with a parametric cavity},\ }\href
  {https://doi.org/10.1103/PhysRevLett.127.100503} {\bibfield  {journal}
  {\bibinfo  {journal} {Physical Review Letters}\ }\textbf {\bibinfo {volume}
  {127}},\ \bibinfo {pages} {100503} (\bibinfo {year} {2021})}\BibitemShut
  {NoStop}%
\bibitem [{\citenamefont {Karamlou}\ \emph {et~al.}(2022)\citenamefont
  {Karamlou}, \citenamefont {Braum\"{u}ller}, \citenamefont {Yanay},
  \citenamefont {Paolo}, \citenamefont {Harrington}, \citenamefont {Kannan},
  \citenamefont {Kim}, \citenamefont {Kjaergaard}, \citenamefont {Melville},
  \citenamefont {Muschinske}, \citenamefont {Niedzielski}, \citenamefont
  {Veps\"{a}l\"{a}inen}, \citenamefont {Winik}, \citenamefont {Yoder},
  \citenamefont {Schwartz}, \citenamefont {Tahan}, \citenamefont {Orlando},
  \citenamefont {Gustavsson},\ and\ \citenamefont {Oliver}}]{Karamlou2022}%
  \BibitemOpen
  \bibfield  {author} {\bibinfo {author} {\bibfnamefont {A.~H.}\ \bibnamefont
  {Karamlou}}, \bibinfo {author} {\bibfnamefont {J.}~\bibnamefont
  {Braum\"{u}ller}}, \bibinfo {author} {\bibfnamefont {Y.}~\bibnamefont
  {Yanay}}, \bibinfo {author} {\bibfnamefont {A.~D.}\ \bibnamefont {Paolo}},
  \bibinfo {author} {\bibfnamefont {P.~M.}\ \bibnamefont {Harrington}},
  \bibinfo {author} {\bibfnamefont {B.}~\bibnamefont {Kannan}}, \bibinfo
  {author} {\bibfnamefont {D.}~\bibnamefont {Kim}}, \bibinfo {author}
  {\bibfnamefont {M.}~\bibnamefont {Kjaergaard}}, \bibinfo {author}
  {\bibfnamefont {A.}~\bibnamefont {Melville}}, \bibinfo {author}
  {\bibfnamefont {S.}~\bibnamefont {Muschinske}}, \bibinfo {author}
  {\bibfnamefont {B.~M.}\ \bibnamefont {Niedzielski}}, \bibinfo {author}
  {\bibfnamefont {A.}~\bibnamefont {Veps\"{a}l\"{a}inen}}, \bibinfo {author}
  {\bibfnamefont {R.}~\bibnamefont {Winik}}, \bibinfo {author} {\bibfnamefont
  {J.~L.}\ \bibnamefont {Yoder}}, \bibinfo {author} {\bibfnamefont
  {M.}~\bibnamefont {Schwartz}}, \bibinfo {author} {\bibfnamefont
  {C.}~\bibnamefont {Tahan}}, \bibinfo {author} {\bibfnamefont {T.~P.}\
  \bibnamefont {Orlando}}, \bibinfo {author} {\bibfnamefont {S.}~\bibnamefont
  {Gustavsson}},\ and\ \bibinfo {author} {\bibfnamefont {W.~D.}\ \bibnamefont
  {Oliver}},\ }\bibfield  {title} {\bibinfo {title} {Quantum transport and
  localization in 1d and 2d tight-binding lattices},\ }\href
  {https://doi.org/10.1038/s41534-022-00528-0} {\bibfield  {journal} {\bibinfo
  {journal} {npj Quantum Information}\ }\textbf {\bibinfo {volume} {8}}
  (\bibinfo {year} {2022})}\BibitemShut {NoStop}%
\bibitem [{\citenamefont {An}\ \emph {et~al.}(2017)\citenamefont {An},
  \citenamefont {Meier},\ and\ \citenamefont {Gadway}}]{An2017}%
  \BibitemOpen
  \bibfield  {author} {\bibinfo {author} {\bibfnamefont {F.~A.}\ \bibnamefont
  {An}}, \bibinfo {author} {\bibfnamefont {E.~J.}\ \bibnamefont {Meier}},\ and\
  \bibinfo {author} {\bibfnamefont {B.}~\bibnamefont {Gadway}},\ }\bibfield
  {title} {\bibinfo {title} {Direct observation of chiral currents and magnetic
  reflection in atomic flux lattices},\ }\href
  {https://doi.org/10.1126/sciadv.1602685} {\bibfield  {journal} {\bibinfo
  {journal} {Science Advances}\ }\textbf {\bibinfo {volume} {3}} (\bibinfo
  {year} {2017})}\BibitemShut {NoStop}%
\bibitem [{\citenamefont {Periwal}\ \emph {et~al.}(2021)\citenamefont
  {Periwal}, \citenamefont {Cooper}, \citenamefont {Kunkel}, \citenamefont
  {Wienand}, \citenamefont {Davis},\ and\ \citenamefont
  {Schleier-Smith}}]{Periwal2021}%
  \BibitemOpen
  \bibfield  {author} {\bibinfo {author} {\bibfnamefont {A.}~\bibnamefont
  {Periwal}}, \bibinfo {author} {\bibfnamefont {E.~S.}\ \bibnamefont {Cooper}},
  \bibinfo {author} {\bibfnamefont {P.}~\bibnamefont {Kunkel}}, \bibinfo
  {author} {\bibfnamefont {J.~F.}\ \bibnamefont {Wienand}}, \bibinfo {author}
  {\bibfnamefont {E.~J.}\ \bibnamefont {Davis}},\ and\ \bibinfo {author}
  {\bibfnamefont {M.}~\bibnamefont {Schleier-Smith}},\ }\bibfield  {title}
  {\bibinfo {title} {Programmable interactions and emergent geometry in an
  array~of atom clouds},\ }\href {https://doi.org/10.1038/s41586-021-04156-0}
  {\bibfield  {journal} {\bibinfo  {journal} {Nature}\ }\textbf {\bibinfo
  {volume} {600}},\ \bibinfo {pages} {630} (\bibinfo {year}
  {2021})}\BibitemShut {NoStop}%
\bibitem [{\citenamefont {Ozawa}\ \emph {et~al.}(2019)\citenamefont {Ozawa},
  \citenamefont {Price}, \citenamefont {Amo}, \citenamefont {Goldman},
  \citenamefont {Hafezi}, \citenamefont {Lu}, \citenamefont {Rechtsman},
  \citenamefont {Schuster}, \citenamefont {Simon}, \citenamefont {Zilberberg},\
  and\ \citenamefont {Carusotto}}]{Ozawa2019}%
  \BibitemOpen
  \bibfield  {author} {\bibinfo {author} {\bibfnamefont {T.}~\bibnamefont
  {Ozawa}}, \bibinfo {author} {\bibfnamefont {H.~M.}\ \bibnamefont {Price}},
  \bibinfo {author} {\bibfnamefont {A.}~\bibnamefont {Amo}}, \bibinfo {author}
  {\bibfnamefont {N.}~\bibnamefont {Goldman}}, \bibinfo {author} {\bibfnamefont
  {M.}~\bibnamefont {Hafezi}}, \bibinfo {author} {\bibfnamefont
  {L.}~\bibnamefont {Lu}}, \bibinfo {author} {\bibfnamefont {M.~C.}\
  \bibnamefont {Rechtsman}}, \bibinfo {author} {\bibfnamefont {D.}~\bibnamefont
  {Schuster}}, \bibinfo {author} {\bibfnamefont {J.}~\bibnamefont {Simon}},
  \bibinfo {author} {\bibfnamefont {O.}~\bibnamefont {Zilberberg}},\ and\
  \bibinfo {author} {\bibfnamefont {I.}~\bibnamefont {Carusotto}},\ }\bibfield
  {title} {\bibinfo {title} {Topological photonics},\ }\href
  {https://doi.org/10.1103/RevModPhys.91.015006} {\bibfield  {journal}
  {\bibinfo  {journal} {Reviews of Modern Physics}\ }\textbf {\bibinfo {volume}
  {91}},\ \bibinfo {pages} {015006} (\bibinfo {year} {2019})}\BibitemShut
  {NoStop}%
\bibitem [{\citenamefont {Lustig}\ \emph {et~al.}(2019)\citenamefont {Lustig},
  \citenamefont {Weimann}, \citenamefont {Plotnik}, \citenamefont {Lumer},
  \citenamefont {Bandres}, \citenamefont {Szameit},\ and\ \citenamefont
  {Segev}}]{Lustig2019}%
  \BibitemOpen
  \bibfield  {author} {\bibinfo {author} {\bibfnamefont {E.}~\bibnamefont
  {Lustig}}, \bibinfo {author} {\bibfnamefont {S.}~\bibnamefont {Weimann}},
  \bibinfo {author} {\bibfnamefont {Y.}~\bibnamefont {Plotnik}}, \bibinfo
  {author} {\bibfnamefont {Y.}~\bibnamefont {Lumer}}, \bibinfo {author}
  {\bibfnamefont {M.~A.}\ \bibnamefont {Bandres}}, \bibinfo {author}
  {\bibfnamefont {A.}~\bibnamefont {Szameit}},\ and\ \bibinfo {author}
  {\bibfnamefont {M.}~\bibnamefont {Segev}},\ }\bibfield  {title} {\bibinfo
  {title} {Photonic topological insulator in synthetic dimensions},\ }\href
  {https://doi.org/10.1038/s41586-019-0943-7} {\bibfield  {journal} {\bibinfo
  {journal} {Nature}\ }\textbf {\bibinfo {volume} {567}},\ \bibinfo {pages}
  {356} (\bibinfo {year} {2019})}\BibitemShut {NoStop}%
\bibitem [{\citenamefont {Yang}\ \emph {et~al.}(2019)\citenamefont {Yang},
  \citenamefont {Peng}, \citenamefont {Zhu}, \citenamefont {Buljan},
  \citenamefont {Joannopoulos}, \citenamefont {Zhen},\ and\ \citenamefont
  {Solja{\v{c}}i{\'c}}}]{Yang2019}%
  \BibitemOpen
  \bibfield  {author} {\bibinfo {author} {\bibfnamefont {Y.}~\bibnamefont
  {Yang}}, \bibinfo {author} {\bibfnamefont {C.}~\bibnamefont {Peng}}, \bibinfo
  {author} {\bibfnamefont {D.}~\bibnamefont {Zhu}}, \bibinfo {author}
  {\bibfnamefont {H.}~\bibnamefont {Buljan}}, \bibinfo {author} {\bibfnamefont
  {J.~D.}\ \bibnamefont {Joannopoulos}}, \bibinfo {author} {\bibfnamefont
  {B.}~\bibnamefont {Zhen}},\ and\ \bibinfo {author} {\bibfnamefont
  {M.}~\bibnamefont {Solja{\v{c}}i{\'c}}},\ }\bibfield  {title} {\bibinfo
  {title} {Synthesis and observation of non-{A}belian gauge fields in real
  space},\ }\href {https://doi.org/10.1126/science.aay3183} {\bibfield
  {journal} {\bibinfo  {journal} {Science}\ }\textbf {\bibinfo {volume}
  {365}},\ \bibinfo {pages} {1021} (\bibinfo {year} {2019})}\BibitemShut
  {NoStop}%
\bibitem [{\citenamefont {Dutt}\ \emph {et~al.}(2020)\citenamefont {Dutt},
  \citenamefont {Lin}, \citenamefont {Yuan}, \citenamefont {Minkov},
  \citenamefont {Xiao},\ and\ \citenamefont {Fan}}]{Dutt2020}%
  \BibitemOpen
  \bibfield  {author} {\bibinfo {author} {\bibfnamefont {A.}~\bibnamefont
  {Dutt}}, \bibinfo {author} {\bibfnamefont {Q.}~\bibnamefont {Lin}}, \bibinfo
  {author} {\bibfnamefont {L.}~\bibnamefont {Yuan}}, \bibinfo {author}
  {\bibfnamefont {M.}~\bibnamefont {Minkov}}, \bibinfo {author} {\bibfnamefont
  {M.}~\bibnamefont {Xiao}},\ and\ \bibinfo {author} {\bibfnamefont
  {S.}~\bibnamefont {Fan}},\ }\bibfield  {title} {\bibinfo {title} {A single
  photonic cavity with two independent physical synthetic dimensions},\ }\href
  {https://doi.org/10.1126/science.aaz3071} {\bibfield  {journal} {\bibinfo
  {journal} {Science}\ }\textbf {\bibinfo {volume} {367}} (\bibinfo {year}
  {2020})}\BibitemShut {NoStop}%
\bibitem [{\citenamefont {Lustig}\ and\ \citenamefont
  {Segev}(2021)}]{Lustig2021}%
  \BibitemOpen
  \bibfield  {author} {\bibinfo {author} {\bibfnamefont {E.}~\bibnamefont
  {Lustig}}\ and\ \bibinfo {author} {\bibfnamefont {M.}~\bibnamefont {Segev}},\
  }\bibfield  {title} {\bibinfo {title} {Topological photonics in synthetic
  dimensions},\ }\href {https://doi.org/10.1364/AOP.418074} {\bibfield
  {journal} {\bibinfo  {journal} {Adv. Opt. Photon.}\ }\textbf {\bibinfo
  {volume} {13}},\ \bibinfo {pages} {426} (\bibinfo {year} {2021})}\BibitemShut
  {NoStop}%
\bibitem [{\citenamefont {Wang}\ \emph
  {et~al.}(2021{\natexlab{a}})\citenamefont {Wang}, \citenamefont {Dutt},
  \citenamefont {Yang}, \citenamefont {Wojcik}, \citenamefont {Vučković},\
  and\ \citenamefont {Fan}}]{Wang2021}%
  \BibitemOpen
  \bibfield  {author} {\bibinfo {author} {\bibfnamefont {K.}~\bibnamefont
  {Wang}}, \bibinfo {author} {\bibfnamefont {A.}~\bibnamefont {Dutt}}, \bibinfo
  {author} {\bibfnamefont {K.~Y.}\ \bibnamefont {Yang}}, \bibinfo {author}
  {\bibfnamefont {C.~C.}\ \bibnamefont {Wojcik}}, \bibinfo {author}
  {\bibfnamefont {J.}~\bibnamefont {Vučković}},\ and\ \bibinfo {author}
  {\bibfnamefont {S.}~\bibnamefont {Fan}},\ }\bibfield  {title} {\bibinfo
  {title} {Generating arbitrary topological windings of a non-{H}ermitian
  band},\ }\href {https://doi.org/10.1126/science.abf6568} {\bibfield
  {journal} {\bibinfo  {journal} {Science}\ }\textbf {\bibinfo {volume}
  {371}},\ \bibinfo {pages} {1240} (\bibinfo {year}
  {2021}{\natexlab{a}})}\BibitemShut {NoStop}%
\bibitem [{\citenamefont {Leefmans}\ \emph {et~al.}(2022)\citenamefont
  {Leefmans}, \citenamefont {Dutt}, \citenamefont {Williams}, \citenamefont
  {Yuan}, \citenamefont {Parto}, \citenamefont {Nori}, \citenamefont {Fan},\
  and\ \citenamefont {Marandi}}]{Leefmans2022}%
  \BibitemOpen
  \bibfield  {author} {\bibinfo {author} {\bibfnamefont {C.}~\bibnamefont
  {Leefmans}}, \bibinfo {author} {\bibfnamefont {A.}~\bibnamefont {Dutt}},
  \bibinfo {author} {\bibfnamefont {J.}~\bibnamefont {Williams}}, \bibinfo
  {author} {\bibfnamefont {L.}~\bibnamefont {Yuan}}, \bibinfo {author}
  {\bibfnamefont {M.}~\bibnamefont {Parto}}, \bibinfo {author} {\bibfnamefont
  {F.}~\bibnamefont {Nori}}, \bibinfo {author} {\bibfnamefont {S.}~\bibnamefont
  {Fan}},\ and\ \bibinfo {author} {\bibfnamefont {A.}~\bibnamefont {Marandi}},\
  }\bibfield  {title} {\bibinfo {title} {Topological dissipation in a
  time-multiplexed photonic resonator network},\ }\href
  {https://doi.org/10.1038/s41567-021-01492-w} {\bibfield  {journal} {\bibinfo
  {journal} {Nature Physics}\ }\textbf {\bibinfo {volume} {18}},\ \bibinfo
  {pages} {442} (\bibinfo {year} {2022})}\BibitemShut {NoStop}%
\bibitem [{\citenamefont {Regensburger}\ \emph {et~al.}(2012)\citenamefont
  {Regensburger}, \citenamefont {Bersch}, \citenamefont {Miri}, \citenamefont
  {Onishchukov}, \citenamefont {Christodoulides},\ and\ \citenamefont
  {Peschel}}]{Regensburger2012}%
  \BibitemOpen
  \bibfield  {author} {\bibinfo {author} {\bibfnamefont {A.}~\bibnamefont
  {Regensburger}}, \bibinfo {author} {\bibfnamefont {C.}~\bibnamefont
  {Bersch}}, \bibinfo {author} {\bibfnamefont {M.-A.}\ \bibnamefont {Miri}},
  \bibinfo {author} {\bibfnamefont {G.}~\bibnamefont {Onishchukov}}, \bibinfo
  {author} {\bibfnamefont {D.~N.}\ \bibnamefont {Christodoulides}},\ and\
  \bibinfo {author} {\bibfnamefont {U.}~\bibnamefont {Peschel}},\ }\bibfield
  {title} {\bibinfo {title} {Parity–time synthetic photonic lattices},\
  }\href {https://doi.org/10.1038/nature11298} {\bibfield  {journal} {\bibinfo
  {journal} {Nature}\ }\textbf {\bibinfo {volume} {488}} (\bibinfo {year}
  {2012})}\BibitemShut {NoStop}%
\bibitem [{\citenamefont {Eichelkraut}\ \emph {et~al.}(2013)\citenamefont
  {Eichelkraut}, \citenamefont {Heilmann}, \citenamefont {Weimann},
  \citenamefont {Stützer}, \citenamefont {Dreisow}, \citenamefont
  {Christodoulides}, \citenamefont {Nolte},\ and\ \citenamefont
  {Szameit}}]{Eichelkraut2013}%
  \BibitemOpen
  \bibfield  {author} {\bibinfo {author} {\bibfnamefont {T.}~\bibnamefont
  {Eichelkraut}}, \bibinfo {author} {\bibfnamefont {R.}~\bibnamefont
  {Heilmann}}, \bibinfo {author} {\bibfnamefont {S.}~\bibnamefont {Weimann}},
  \bibinfo {author} {\bibfnamefont {S.}~\bibnamefont {Stützer}}, \bibinfo
  {author} {\bibfnamefont {F.}~\bibnamefont {Dreisow}}, \bibinfo {author}
  {\bibfnamefont {D.~N.}\ \bibnamefont {Christodoulides}}, \bibinfo {author}
  {\bibfnamefont {S.}~\bibnamefont {Nolte}},\ and\ \bibinfo {author}
  {\bibfnamefont {A.}~\bibnamefont {Szameit}},\ }\bibfield  {title} {\bibinfo
  {title} {Mobility transition from ballistic to diffusive transport in
  non-{H}ermitian lattices},\ }\href {https://doi.org/10.1038/ncomms3533}
  {\bibfield  {journal} {\bibinfo  {journal} {Nature Communications}\ }\textbf
  {\bibinfo {volume} {4}} (\bibinfo {year} {2013})}\BibitemShut {NoStop}%
\bibitem [{\citenamefont {Hodaei}\ \emph {et~al.}(2017)\citenamefont {Hodaei},
  \citenamefont {Hassan}, \citenamefont {Wittek}, \citenamefont
  {Garcia-Gracia}, \citenamefont {El-Ganainy}, \citenamefont
  {Christodoulides},\ and\ \citenamefont {Khajavikhan}}]{Hodaei2017}%
  \BibitemOpen
  \bibfield  {author} {\bibinfo {author} {\bibfnamefont {H.}~\bibnamefont
  {Hodaei}}, \bibinfo {author} {\bibfnamefont {A.~U.}\ \bibnamefont {Hassan}},
  \bibinfo {author} {\bibfnamefont {S.}~\bibnamefont {Wittek}}, \bibinfo
  {author} {\bibfnamefont {H.}~\bibnamefont {Garcia-Gracia}}, \bibinfo {author}
  {\bibfnamefont {R.}~\bibnamefont {El-Ganainy}}, \bibinfo {author}
  {\bibfnamefont {D.~N.}\ \bibnamefont {Christodoulides}},\ and\ \bibinfo
  {author} {\bibfnamefont {M.}~\bibnamefont {Khajavikhan}},\ }\bibfield
  {title} {\bibinfo {title} {Enhanced sensitivity at higher-order exceptional
  points},\ }\href {https://doi.org/10.1038/nature23280} {\bibfield  {journal}
  {\bibinfo  {journal} {Nature}\ }\textbf {\bibinfo {volume} {548}},\ \bibinfo
  {pages} {187} (\bibinfo {year} {2017})}\BibitemShut {NoStop}%
\bibitem [{\citenamefont {Weidemann}\ \emph {et~al.}(2021)\citenamefont
  {Weidemann}, \citenamefont {Kremer}, \citenamefont {Longhi},\ and\
  \citenamefont {Szameit}}]{Weidemann2021}%
  \BibitemOpen
  \bibfield  {author} {\bibinfo {author} {\bibfnamefont {S.}~\bibnamefont
  {Weidemann}}, \bibinfo {author} {\bibfnamefont {M.}~\bibnamefont {Kremer}},
  \bibinfo {author} {\bibfnamefont {S.}~\bibnamefont {Longhi}},\ and\ \bibinfo
  {author} {\bibfnamefont {A.}~\bibnamefont {Szameit}},\ }\bibfield  {title}
  {\bibinfo {title} {Coexistence of dynamical delocalization and spectral
  localization through stochastic dissipation},\ }\href
  {https://doi.org/10.1038/s41566-021-00823-w} {\bibfield  {journal} {\bibinfo
  {journal} {Nature Photonics}\ }\textbf {\bibinfo {volume} {15}} (\bibinfo
  {year} {2021})}\BibitemShut {NoStop}%
\bibitem [{\citenamefont {Xia}\ \emph {et~al.}(2021)\citenamefont {Xia},
  \citenamefont {Kaltsas}, \citenamefont {Song}, \citenamefont {Komis},
  \citenamefont {Xu}, \citenamefont {Szameit}, \citenamefont {Buljan},
  \citenamefont {Makris},\ and\ \citenamefont {Chen}}]{Xia2021}%
  \BibitemOpen
  \bibfield  {author} {\bibinfo {author} {\bibfnamefont {S.}~\bibnamefont
  {Xia}}, \bibinfo {author} {\bibfnamefont {D.}~\bibnamefont {Kaltsas}},
  \bibinfo {author} {\bibfnamefont {D.}~\bibnamefont {Song}}, \bibinfo {author}
  {\bibfnamefont {I.}~\bibnamefont {Komis}}, \bibinfo {author} {\bibfnamefont
  {J.}~\bibnamefont {Xu}}, \bibinfo {author} {\bibfnamefont {A.}~\bibnamefont
  {Szameit}}, \bibinfo {author} {\bibfnamefont {H.}~\bibnamefont {Buljan}},
  \bibinfo {author} {\bibfnamefont {K.~G.}\ \bibnamefont {Makris}},\ and\
  \bibinfo {author} {\bibfnamefont {Z.}~\bibnamefont {Chen}},\ }\bibfield
  {title} {\bibinfo {title} {Nonlinear tuning of {PT} symmetry and
  {non-Hermitian} topological states},\ }\href
  {https://doi.org/10.1126/science.abf6873} {\bibfield  {journal} {\bibinfo
  {journal} {Science}\ }\textbf {\bibinfo {volume} {372}},\ \bibinfo {pages}
  {72} (\bibinfo {year} {2021})}\BibitemShut {NoStop}%
\bibitem [{\citenamefont {Bandres}\ \emph {et~al.}(2018)\citenamefont
  {Bandres}, \citenamefont {Wittek}, \citenamefont {Harari}, \citenamefont
  {Parto}, \citenamefont {Ren}, \citenamefont {Segev}, \citenamefont
  {Christodoulides},\ and\ \citenamefont {Khajavikhan}}]{Bandres2018}%
  \BibitemOpen
  \bibfield  {author} {\bibinfo {author} {\bibfnamefont {M.~A.}\ \bibnamefont
  {Bandres}}, \bibinfo {author} {\bibfnamefont {S.}~\bibnamefont {Wittek}},
  \bibinfo {author} {\bibfnamefont {G.}~\bibnamefont {Harari}}, \bibinfo
  {author} {\bibfnamefont {M.}~\bibnamefont {Parto}}, \bibinfo {author}
  {\bibfnamefont {J.}~\bibnamefont {Ren}}, \bibinfo {author} {\bibfnamefont
  {M.}~\bibnamefont {Segev}}, \bibinfo {author} {\bibfnamefont {D.~N.}\
  \bibnamefont {Christodoulides}},\ and\ \bibinfo {author} {\bibfnamefont
  {M.}~\bibnamefont {Khajavikhan}},\ }\bibfield  {title} {\bibinfo {title}
  {Topological insulator laser: Experiments},\ }\href
  {https://doi.org/10.1126/science.aar4005} {\bibfield  {journal} {\bibinfo
  {journal} {Science}\ }\textbf {\bibinfo {volume} {359}} (\bibinfo {year}
  {2018})}\BibitemShut {NoStop}%
\bibitem [{\citenamefont {Solnyshkov}\ \emph {et~al.}(2018)\citenamefont
  {Solnyshkov}, \citenamefont {Bleu},\ and\ \citenamefont
  {Malpuech}}]{Solnyshkov2018}%
  \BibitemOpen
  \bibfield  {author} {\bibinfo {author} {\bibfnamefont {D.~D.}\ \bibnamefont
  {Solnyshkov}}, \bibinfo {author} {\bibfnamefont {O.}~\bibnamefont {Bleu}},\
  and\ \bibinfo {author} {\bibfnamefont {G.}~\bibnamefont {Malpuech}},\
  }\bibfield  {title} {\bibinfo {title} {Topological optical isolator based on
  polariton graphene},\ }\href {https://doi.org/10.1063/1.5018902} {\bibfield
  {journal} {\bibinfo  {journal} {Applied Physics Letters}\ }\textbf {\bibinfo
  {volume} {112}},\ \bibinfo {pages} {031106} (\bibinfo {year}
  {2018})}\BibitemShut {NoStop}%
\bibitem [{\citenamefont {Hokmabadi}\ \emph {et~al.}(2019)\citenamefont
  {Hokmabadi}, \citenamefont {Schumer}, \citenamefont {Christodoulides},\ and\
  \citenamefont {Khajavikhan}}]{Hokmabadi2019}%
  \BibitemOpen
  \bibfield  {author} {\bibinfo {author} {\bibfnamefont {M.~P.}\ \bibnamefont
  {Hokmabadi}}, \bibinfo {author} {\bibfnamefont {A.}~\bibnamefont {Schumer}},
  \bibinfo {author} {\bibfnamefont {D.~N.}\ \bibnamefont {Christodoulides}},\
  and\ \bibinfo {author} {\bibfnamefont {M.}~\bibnamefont {Khajavikhan}},\
  }\bibfield  {title} {\bibinfo {title} {Non-hermitian ring laser gyroscopes
  with enhanced sagnac sensitivity},\ }\href
  {https://doi.org/10.1038/s41586-019-1780-4} {\bibfield  {journal} {\bibinfo
  {journal} {Nature}\ }\textbf {\bibinfo {volume} {576}},\ \bibinfo {pages}
  {70} (\bibinfo {year} {2019})}\BibitemShut {NoStop}%
\bibitem [{\citenamefont {Zeng}\ \emph {et~al.}(2020)\citenamefont {Zeng},
  \citenamefont {Chattopadhyay}, \citenamefont {Zhu}, \citenamefont {Qiang},
  \citenamefont {Li}, \citenamefont {Jin}, \citenamefont {Li}, \citenamefont
  {Davies}, \citenamefont {Linfield}, \citenamefont {Zhang}, \citenamefont
  {Chong},\ and\ \citenamefont {Wang}}]{Zeng2020}%
  \BibitemOpen
  \bibfield  {author} {\bibinfo {author} {\bibfnamefont {Y.}~\bibnamefont
  {Zeng}}, \bibinfo {author} {\bibfnamefont {U.}~\bibnamefont {Chattopadhyay}},
  \bibinfo {author} {\bibfnamefont {B.}~\bibnamefont {Zhu}}, \bibinfo {author}
  {\bibfnamefont {B.}~\bibnamefont {Qiang}}, \bibinfo {author} {\bibfnamefont
  {J.}~\bibnamefont {Li}}, \bibinfo {author} {\bibfnamefont {Y.}~\bibnamefont
  {Jin}}, \bibinfo {author} {\bibfnamefont {L.}~\bibnamefont {Li}}, \bibinfo
  {author} {\bibfnamefont {A.~G.}\ \bibnamefont {Davies}}, \bibinfo {author}
  {\bibfnamefont {E.~H.}\ \bibnamefont {Linfield}}, \bibinfo {author}
  {\bibfnamefont {B.}~\bibnamefont {Zhang}}, \bibinfo {author} {\bibfnamefont
  {Y.}~\bibnamefont {Chong}},\ and\ \bibinfo {author} {\bibfnamefont {Q.~J.}\
  \bibnamefont {Wang}},\ }\bibfield  {title} {\bibinfo {title} {Electrically
  pumped topological laser with valley edge modes},\ }\href
  {https://doi.org/10.1038/s41586-020-1981-x} {\bibfield  {journal} {\bibinfo
  {journal} {Nature}\ }\textbf {\bibinfo {volume} {578}},\ \bibinfo {pages}
  {246} (\bibinfo {year} {2020})}\BibitemShut {NoStop}%
\bibitem [{\citenamefont {Bersch}\ \emph {et~al.}(2009)\citenamefont {Bersch},
  \citenamefont {Onishchukov},\ and\ \citenamefont {Peschel}}]{Bersch2009}%
  \BibitemOpen
  \bibfield  {author} {\bibinfo {author} {\bibfnamefont {C.}~\bibnamefont
  {Bersch}}, \bibinfo {author} {\bibfnamefont {G.}~\bibnamefont
  {Onishchukov}},\ and\ \bibinfo {author} {\bibfnamefont {U.}~\bibnamefont
  {Peschel}},\ }\bibfield  {title} {\bibinfo {title} {Experimental observation
  of spectral {B}loch oscillations},\ }\href
  {https://doi.org/10.1364/OL.34.002372} {\bibfield  {journal} {\bibinfo
  {journal} {Optics Letters}\ }\textbf {\bibinfo {volume} {34}} (\bibinfo
  {year} {2009})}\BibitemShut {NoStop}%
\bibitem [{\citenamefont {Schwartz}\ and\ \citenamefont
  {Fischer}(2013)}]{Schwartz2013}%
  \BibitemOpen
  \bibfield  {author} {\bibinfo {author} {\bibfnamefont {A.}~\bibnamefont
  {Schwartz}}\ and\ \bibinfo {author} {\bibfnamefont {B.}~\bibnamefont
  {Fischer}},\ }\bibfield  {title} {\bibinfo {title} {Laser mode hyper-combs},\
  }\href {https://doi.org/10.1364/OE.21.006196} {\bibfield  {journal} {\bibinfo
   {journal} {Optics Express}\ }\textbf {\bibinfo {volume} {21}},\ \bibinfo
  {pages} {6196} (\bibinfo {year} {2013})}\BibitemShut {NoStop}%
\bibitem [{\citenamefont {Ozawa}\ \emph {et~al.}(2016)\citenamefont {Ozawa},
  \citenamefont {Price}, \citenamefont {Goldman}, \citenamefont {Zilberberg},\
  and\ \citenamefont {Carusotto}}]{Ozawa2016}%
  \BibitemOpen
  \bibfield  {author} {\bibinfo {author} {\bibfnamefont {T.}~\bibnamefont
  {Ozawa}}, \bibinfo {author} {\bibfnamefont {H.~M.}\ \bibnamefont {Price}},
  \bibinfo {author} {\bibfnamefont {N.}~\bibnamefont {Goldman}}, \bibinfo
  {author} {\bibfnamefont {O.}~\bibnamefont {Zilberberg}},\ and\ \bibinfo
  {author} {\bibfnamefont {I.}~\bibnamefont {Carusotto}},\ }\bibfield  {title}
  {\bibinfo {title} {Synthetic dimensions in integrated photonics: From optical
  isolation to four-dimensional quantum {H}all physics},\ }\href
  {https://doi.org/10.1103/PhysRevA.93.043827} {\bibfield  {journal} {\bibinfo
  {journal} {Physical Review A}\ }\textbf {\bibinfo {volume} {93}},\ \bibinfo
  {pages} {043827} (\bibinfo {year} {2016})}\BibitemShut {NoStop}%
\bibitem [{\citenamefont {Yuan}\ and\ \citenamefont {Fan}(2016)}]{Yuan2016}%
  \BibitemOpen
  \bibfield  {author} {\bibinfo {author} {\bibfnamefont {L.}~\bibnamefont
  {Yuan}}\ and\ \bibinfo {author} {\bibfnamefont {S.}~\bibnamefont {Fan}},\
  }\bibfield  {title} {\bibinfo {title} {Bloch oscillation and unidirectional
  translation of frequency in a dynamically modulated ring resonator},\ }\href
  {https://doi.org/10.1364/OPTICA.3.001014} {\bibfield  {journal} {\bibinfo
  {journal} {Optica}\ }\textbf {\bibinfo {volume} {3}} (\bibinfo {year}
  {2016})}\BibitemShut {NoStop}%
\bibitem [{\citenamefont {Bell}\ \emph {et~al.}(2017)\citenamefont {Bell},
  \citenamefont {Wang}, \citenamefont {Solntsev}, \citenamefont {Neshev},
  \citenamefont {Sukhorukov},\ and\ \citenamefont {Eggleton}}]{Bell2017}%
  \BibitemOpen
  \bibfield  {author} {\bibinfo {author} {\bibfnamefont {B.~A.}\ \bibnamefont
  {Bell}}, \bibinfo {author} {\bibfnamefont {K.}~\bibnamefont {Wang}}, \bibinfo
  {author} {\bibfnamefont {A.~S.}\ \bibnamefont {Solntsev}}, \bibinfo {author}
  {\bibfnamefont {D.~N.}\ \bibnamefont {Neshev}}, \bibinfo {author}
  {\bibfnamefont {A.~A.}\ \bibnamefont {Sukhorukov}},\ and\ \bibinfo {author}
  {\bibfnamefont {B.~J.}\ \bibnamefont {Eggleton}},\ }\bibfield  {title}
  {\bibinfo {title} {Spectral photonic lattices with complex long-range
  coupling},\ }\href {https://doi.org/10.1364/OPTICA.4.001433} {\bibfield
  {journal} {\bibinfo  {journal} {Optica}\ }\textbf {\bibinfo {volume} {4}}
  (\bibinfo {year} {2017})}\BibitemShut {NoStop}%
\bibitem [{\citenamefont {Qin}\ \emph {et~al.}(2018{\natexlab{a}})\citenamefont
  {Qin}, \citenamefont {Zhou}, \citenamefont {Peng}, \citenamefont {Sounas},
  \citenamefont {Zhu}, \citenamefont {Wang}, \citenamefont {Dong},
  \citenamefont {Zhang}, \citenamefont {Alù},\ and\ \citenamefont
  {Lu}}]{Qin2018}%
  \BibitemOpen
  \bibfield  {author} {\bibinfo {author} {\bibfnamefont {C.}~\bibnamefont
  {Qin}}, \bibinfo {author} {\bibfnamefont {F.}~\bibnamefont {Zhou}}, \bibinfo
  {author} {\bibfnamefont {Y.}~\bibnamefont {Peng}}, \bibinfo {author}
  {\bibfnamefont {D.}~\bibnamefont {Sounas}}, \bibinfo {author} {\bibfnamefont
  {X.}~\bibnamefont {Zhu}}, \bibinfo {author} {\bibfnamefont {B.}~\bibnamefont
  {Wang}}, \bibinfo {author} {\bibfnamefont {J.}~\bibnamefont {Dong}}, \bibinfo
  {author} {\bibfnamefont {X.}~\bibnamefont {Zhang}}, \bibinfo {author}
  {\bibfnamefont {A.}~\bibnamefont {Alù}},\ and\ \bibinfo {author}
  {\bibfnamefont {P.}~\bibnamefont {Lu}},\ }\bibfield  {title} {\bibinfo
  {title} {Spectrum control through discrete frequency diffraction in the
  presence of photonic gauge potentials},\ }\href
  {https://doi.org/10.1103/PhysRevLett.120.133901} {\bibfield  {journal}
  {\bibinfo  {journal} {Physical Review Letters}\ }\textbf {\bibinfo {volume}
  {120}} (\bibinfo {year} {2018}{\natexlab{a}})}\BibitemShut {NoStop}%
\bibitem [{\citenamefont {Dutt}\ \emph {et~al.}(2019)\citenamefont {Dutt},
  \citenamefont {Minkov}, \citenamefont {Lin}, \citenamefont {Yuan},
  \citenamefont {Miller},\ and\ \citenamefont {Fan}}]{Dutt2019}%
  \BibitemOpen
  \bibfield  {author} {\bibinfo {author} {\bibfnamefont {A.}~\bibnamefont
  {Dutt}}, \bibinfo {author} {\bibfnamefont {M.}~\bibnamefont {Minkov}},
  \bibinfo {author} {\bibfnamefont {Q.}~\bibnamefont {Lin}}, \bibinfo {author}
  {\bibfnamefont {L.}~\bibnamefont {Yuan}}, \bibinfo {author} {\bibfnamefont
  {D.~A.~B.}\ \bibnamefont {Miller}},\ and\ \bibinfo {author} {\bibfnamefont
  {S.}~\bibnamefont {Fan}},\ }\bibfield  {title} {\bibinfo {title}
  {Experimental band structure spectroscopy along a synthetic dimension},\
  }\href {https://doi.org/10.1038/s41467-019-11117-9} {\bibfield  {journal}
  {\bibinfo  {journal} {Nature Communications}\ }\textbf {\bibinfo {volume}
  {10}} (\bibinfo {year} {2019})}\BibitemShut {NoStop}%
\bibitem [{\citenamefont {Hu}\ \emph {et~al.}(2020)\citenamefont {Hu},
  \citenamefont {Reimer}, \citenamefont {Shams-Ansari}, \citenamefont {Zhang},\
  and\ \citenamefont {Lončar}}]{Hu2020}%
  \BibitemOpen
  \bibfield  {author} {\bibinfo {author} {\bibfnamefont {Y.}~\bibnamefont
  {Hu}}, \bibinfo {author} {\bibfnamefont {C.}~\bibnamefont {Reimer}}, \bibinfo
  {author} {\bibfnamefont {A.}~\bibnamefont {Shams-Ansari}}, \bibinfo {author}
  {\bibfnamefont {M.}~\bibnamefont {Zhang}},\ and\ \bibinfo {author}
  {\bibfnamefont {M.}~\bibnamefont {Lončar}},\ }\bibfield  {title} {\bibinfo
  {title} {Realization of high-dimensional frequency crystals in electro-optic
  microcombs},\ }\href {https://doi.org/https://doi.org/10.1364/OPTICA.395114}
  {\bibfield  {journal} {\bibinfo  {journal} {Optica}\ }\textbf {\bibinfo
  {volume} {7}},\ \bibinfo {pages} {1189} (\bibinfo {year} {2020})}\BibitemShut
  {NoStop}%
\bibitem [{\citenamefont {Li}\ \emph {et~al.}(2021)\citenamefont {Li},
  \citenamefont {Zheng}, \citenamefont {Dutt}, \citenamefont {Yu},
  \citenamefont {Shan}, \citenamefont {Liu}, \citenamefont {Yuan},
  \citenamefont {Fan},\ and\ \citenamefont {Chen}}]{Li2021}%
  \BibitemOpen
  \bibfield  {author} {\bibinfo {author} {\bibfnamefont {G.}~\bibnamefont
  {Li}}, \bibinfo {author} {\bibfnamefont {Y.}~\bibnamefont {Zheng}}, \bibinfo
  {author} {\bibfnamefont {A.}~\bibnamefont {Dutt}}, \bibinfo {author}
  {\bibfnamefont {D.}~\bibnamefont {Yu}}, \bibinfo {author} {\bibfnamefont
  {Q.}~\bibnamefont {Shan}}, \bibinfo {author} {\bibfnamefont {S.}~\bibnamefont
  {Liu}}, \bibinfo {author} {\bibfnamefont {L.}~\bibnamefont {Yuan}}, \bibinfo
  {author} {\bibfnamefont {S.}~\bibnamefont {Fan}},\ and\ \bibinfo {author}
  {\bibfnamefont {X.}~\bibnamefont {Chen}},\ }\bibfield  {title} {\bibinfo
  {title} {Dynamic band structure measurement in the synthetic space},\ }\href
  {https://doi.org/10.1126/sciadv.abe4335} {\bibfield  {journal} {\bibinfo
  {journal} {Science Advances}\ }\textbf {\bibinfo {volume} {7}} (\bibinfo
  {year} {2021})}\BibitemShut {NoStop}%
\bibitem [{\citenamefont {Chen}\ \emph {et~al.}(2021)\citenamefont {Chen},
  \citenamefont {Yang}, \citenamefont {Qin}, \citenamefont {Li}, \citenamefont
  {Wang}, \citenamefont {Han}, \citenamefont {Zhang}, \citenamefont {Liu},
  \citenamefont {Wang}, \citenamefont {Long}, \citenamefont {Zhang},\ and\
  \citenamefont {Lu}}]{Chen2021}%
  \BibitemOpen
  \bibfield  {author} {\bibinfo {author} {\bibfnamefont {H.}~\bibnamefont
  {Chen}}, \bibinfo {author} {\bibfnamefont {N.~N.}\ \bibnamefont {Yang}},
  \bibinfo {author} {\bibfnamefont {C.}~\bibnamefont {Qin}}, \bibinfo {author}
  {\bibfnamefont {W.}~\bibnamefont {Li}}, \bibinfo {author} {\bibfnamefont
  {B.}~\bibnamefont {Wang}}, \bibinfo {author} {\bibfnamefont {T.}~\bibnamefont
  {Han}}, \bibinfo {author} {\bibfnamefont {C.}~\bibnamefont {Zhang}}, \bibinfo
  {author} {\bibfnamefont {W.}~\bibnamefont {Liu}}, \bibinfo {author}
  {\bibfnamefont {K.}~\bibnamefont {Wang}}, \bibinfo {author} {\bibfnamefont
  {H.}~\bibnamefont {Long}}, \bibinfo {author} {\bibfnamefont {X.}~\bibnamefont
  {Zhang}},\ and\ \bibinfo {author} {\bibfnamefont {P.}~\bibnamefont {Lu}},\
  }\bibfield  {title} {\bibinfo {title} {Real-time observation of frequency
  {B}loch oscillations with fibre loop modulation},\ }\href
  {https://doi.org/10.1038/s41377-021-00494-w} {\bibfield  {journal} {\bibinfo
  {journal} {Light: Science \& Applications}\ }\textbf {\bibinfo {volume} {10}}
  (\bibinfo {year} {2021})}\BibitemShut {NoStop}%
\bibitem [{\citenamefont {Yuan}\ \emph {et~al.}(2021)\citenamefont {Yuan},
  \citenamefont {Dutt},\ and\ \citenamefont {Fan}}]{YuanDutt2021}%
  \BibitemOpen
  \bibfield  {author} {\bibinfo {author} {\bibfnamefont {L.}~\bibnamefont
  {Yuan}}, \bibinfo {author} {\bibfnamefont {A.}~\bibnamefont {Dutt}},\ and\
  \bibinfo {author} {\bibfnamefont {S.}~\bibnamefont {Fan}},\ }\bibfield
  {title} {\bibinfo {title} {Synthetic frequency dimensions in dynamically
  modulated ring resonators},\ }\href {https://doi.org/10.1063/5.0056359}
  {\bibfield  {journal} {\bibinfo  {journal} {APL Photonics}\ }\textbf
  {\bibinfo {volume} {6}} (\bibinfo {year} {2021})}\BibitemShut {NoStop}%
\bibitem [{\citenamefont {Zhao}\ \emph {et~al.}(2021)\citenamefont {Zhao},
  \citenamefont {Li}, \citenamefont {Li},\ and\ \citenamefont {Li}}]{Zhao2021}%
  \BibitemOpen
  \bibfield  {author} {\bibinfo {author} {\bibfnamefont {H.}~\bibnamefont
  {Zhao}}, \bibinfo {author} {\bibfnamefont {B.}~\bibnamefont {Li}}, \bibinfo
  {author} {\bibfnamefont {H.}~\bibnamefont {Li}},\ and\ \bibinfo {author}
  {\bibfnamefont {M.}~\bibnamefont {Li}},\ }\bibfield  {title} {\bibinfo
  {title} {Scaling optical computing in synthetic frequency dimension using
  integrated cavity acousto-optics},\ }\href
  {https://doi.org/10.48550/arXiv.2106.08494} {\bibfield  {journal} {\bibinfo
  {journal} {arXiv preprint arXiv:2106.08494}\ } (\bibinfo {year}
  {2021})}\BibitemShut {NoStop}%
\bibitem [{\citenamefont {Balčytis}\ \emph {et~al.}(2022)\citenamefont
  {Balčytis}, \citenamefont {Ozawa}, \citenamefont {Ota}, \citenamefont
  {Iwamoto}, \citenamefont {Maeda},\ and\ \citenamefont {Baba}}]{Balcytis2022}%
  \BibitemOpen
  \bibfield  {author} {\bibinfo {author} {\bibfnamefont {A.}~\bibnamefont
  {Balčytis}}, \bibinfo {author} {\bibfnamefont {T.}~\bibnamefont {Ozawa}},
  \bibinfo {author} {\bibfnamefont {Y.}~\bibnamefont {Ota}}, \bibinfo {author}
  {\bibfnamefont {S.}~\bibnamefont {Iwamoto}}, \bibinfo {author} {\bibfnamefont
  {J.}~\bibnamefont {Maeda}},\ and\ \bibinfo {author} {\bibfnamefont
  {T.}~\bibnamefont {Baba}},\ }\bibfield  {title} {\bibinfo {title} {Synthetic
  dimension band structures on a {Si CMOS} photonic platform},\ }\href
  {https://doi.org/10.1126/sciadv.abk0468} {\bibfield  {journal} {\bibinfo
  {journal} {Science Advances}\ }\textbf {\bibinfo {volume} {8}} (\bibinfo
  {year} {2022})}\BibitemShut {NoStop}%
\bibitem [{\citenamefont {Dutt}\ \emph {et~al.}(2022)\citenamefont {Dutt},
  \citenamefont {Yuan}, \citenamefont {Yang}, \citenamefont {Wang},
  \citenamefont {Buddhiraju}, \citenamefont {Vučković},\ and\ \citenamefont
  {Fan}}]{Dutt2022}%
  \BibitemOpen
  \bibfield  {author} {\bibinfo {author} {\bibfnamefont {A.}~\bibnamefont
  {Dutt}}, \bibinfo {author} {\bibfnamefont {L.}~\bibnamefont {Yuan}}, \bibinfo
  {author} {\bibfnamefont {K.~Y.}\ \bibnamefont {Yang}}, \bibinfo {author}
  {\bibfnamefont {K.}~\bibnamefont {Wang}}, \bibinfo {author} {\bibfnamefont
  {S.}~\bibnamefont {Buddhiraju}}, \bibinfo {author} {\bibfnamefont
  {J.}~\bibnamefont {Vučković}},\ and\ \bibinfo {author} {\bibfnamefont
  {S.}~\bibnamefont {Fan}},\ }\bibfield  {title} {\bibinfo {title} {Creating
  boundaries along a synthetic frequency dimension},\ }\href
  {https://doi.org/10.1038/s41467-022-31140-7} {\bibfield  {journal} {\bibinfo
  {journal} {Nature Communications}\ }\textbf {\bibinfo {volume} {13}},\
  \bibinfo {pages} {3377} (\bibinfo {year} {2022})}\BibitemShut {NoStop}%
\bibitem [{\citenamefont {Miyake}\ \emph {et~al.}(2013)\citenamefont {Miyake},
  \citenamefont {Siviloglou}, \citenamefont {Kennedy}, \citenamefont {Burton},\
  and\ \citenamefont {Ketterle}}]{Miyake2013}%
  \BibitemOpen
  \bibfield  {author} {\bibinfo {author} {\bibfnamefont {H.}~\bibnamefont
  {Miyake}}, \bibinfo {author} {\bibfnamefont {G.~A.}\ \bibnamefont
  {Siviloglou}}, \bibinfo {author} {\bibfnamefont {C.~J.}\ \bibnamefont
  {Kennedy}}, \bibinfo {author} {\bibfnamefont {W.~C.}\ \bibnamefont
  {Burton}},\ and\ \bibinfo {author} {\bibfnamefont {W.}~\bibnamefont
  {Ketterle}},\ }\bibfield  {title} {\bibinfo {title} {Realizing the {H}arper
  {H}amiltonian with laser-assisted tunneling in optical lattices},\ }\href
  {https://doi.org/10.1103/PhysRevLett.111.185302} {\bibfield  {journal}
  {\bibinfo  {journal} {Physical Review Letters}\ }\textbf {\bibinfo {volume}
  {111}} (\bibinfo {year} {2013})}\BibitemShut {NoStop}%
\bibitem [{\citenamefont {Yuan}\ and\ \citenamefont {Fan}(2015)}]{Yuan2015}%
  \BibitemOpen
  \bibfield  {author} {\bibinfo {author} {\bibfnamefont {L.}~\bibnamefont
  {Yuan}}\ and\ \bibinfo {author} {\bibfnamefont {S.}~\bibnamefont {Fan}},\
  }\bibfield  {title} {\bibinfo {title} {Three-dimensional dynamic localization
  of light from a time-dependent effective gauge field for photons},\ }\href
  {https://doi.org/10.1103/PhysRevLett.114.243901} {\bibfield  {journal}
  {\bibinfo  {journal} {Physical Review Letters}\ }\textbf {\bibinfo {volume}
  {114}} (\bibinfo {year} {2015})}\BibitemShut {NoStop}%
\bibitem [{\citenamefont {Qin}\ \emph {et~al.}(2018{\natexlab{b}})\citenamefont
  {Qin}, \citenamefont {Yuan}, \citenamefont {Wang}, \citenamefont {Fan},\ and\
  \citenamefont {Lu}}]{QinYuan2018}%
  \BibitemOpen
  \bibfield  {author} {\bibinfo {author} {\bibfnamefont {C.}~\bibnamefont
  {Qin}}, \bibinfo {author} {\bibfnamefont {L.}~\bibnamefont {Yuan}}, \bibinfo
  {author} {\bibfnamefont {B.}~\bibnamefont {Wang}}, \bibinfo {author}
  {\bibfnamefont {S.}~\bibnamefont {Fan}},\ and\ \bibinfo {author}
  {\bibfnamefont {P.}~\bibnamefont {Lu}},\ }\bibfield  {title} {\bibinfo
  {title} {Effective electric-field force for a photon in a synthetic frequency
  lattice created in a waveguide modulator},\ }\href
  {https://doi.org/10.1103/PhysRevA.97.063838} {\bibfield  {journal} {\bibinfo
  {journal} {Physical Review A}\ }\textbf {\bibinfo {volume} {97}} (\bibinfo
  {year} {2018}{\natexlab{b}})}\BibitemShut {NoStop}%
\bibitem [{\citenamefont {Peterson}\ \emph {et~al.}(2019)\citenamefont
  {Peterson}, \citenamefont {Benalcazar}, \citenamefont {Lin}, \citenamefont
  {Hughes},\ and\ \citenamefont {Bahl}}]{Peterson2019}%
  \BibitemOpen
  \bibfield  {author} {\bibinfo {author} {\bibfnamefont {C.~W.}\ \bibnamefont
  {Peterson}}, \bibinfo {author} {\bibfnamefont {W.~A.}\ \bibnamefont
  {Benalcazar}}, \bibinfo {author} {\bibfnamefont {M.}~\bibnamefont {Lin}},
  \bibinfo {author} {\bibfnamefont {T.~L.}\ \bibnamefont {Hughes}},\ and\
  \bibinfo {author} {\bibfnamefont {G.}~\bibnamefont {Bahl}},\ }\bibfield
  {title} {\bibinfo {title} {Strong nonreciprocity in modulated resonator
  chains through synthetic electric and magnetic fields},\ }\href
  {https://doi.org/10.1103/physrevlett.123.063901} {\bibfield  {journal}
  {\bibinfo  {journal} {Physical Review Letters}\ }\textbf {\bibinfo {volume}
  {123}} (\bibinfo {year} {2019})}\BibitemShut {NoStop}%
\bibitem [{\citenamefont {Lee}\ \emph {et~al.}(2020)\citenamefont {Lee},
  \citenamefont {Pechal}, \citenamefont {Wollack}, \citenamefont
  {Arrangoiz-Arriola}, \citenamefont {Wang},\ and\ \citenamefont
  {Safavi-Naeni}}]{Lee2020}%
  \BibitemOpen
  \bibfield  {author} {\bibinfo {author} {\bibfnamefont {N.~R.~A.}\
  \bibnamefont {Lee}}, \bibinfo {author} {\bibfnamefont {M.}~\bibnamefont
  {Pechal}}, \bibinfo {author} {\bibfnamefont {E.~A.}\ \bibnamefont {Wollack}},
  \bibinfo {author} {\bibfnamefont {P.}~\bibnamefont {Arrangoiz-Arriola}},
  \bibinfo {author} {\bibfnamefont {Z.}~\bibnamefont {Wang}},\ and\ \bibinfo
  {author} {\bibfnamefont {A.~H.}\ \bibnamefont {Safavi-Naeni}},\ }\bibfield
  {title} {\bibinfo {title} {Propagation of microwave photons along a synthetic
  dimension},\ }\href {https://doi.org/10.1103/PhysRevA.101.053807} {\bibfield
  {journal} {\bibinfo  {journal} {Physical Review A}\ }\textbf {\bibinfo
  {volume} {101}} (\bibinfo {year} {2020})}\BibitemShut {NoStop}%
\bibitem [{\citenamefont {D’Errico}\ \emph {et~al.}(2021)\citenamefont
  {D’Errico}, \citenamefont {Barboza}, \citenamefont {Tudor}, \citenamefont
  {Dauphin}, \citenamefont {Massignan}, \citenamefont {Marrucci},\ and\
  \citenamefont {Cardano}}]{DErrico2021}%
  \BibitemOpen
  \bibfield  {author} {\bibinfo {author} {\bibfnamefont {A.}~\bibnamefont
  {D’Errico}}, \bibinfo {author} {\bibfnamefont {R.}~\bibnamefont {Barboza}},
  \bibinfo {author} {\bibfnamefont {R.}~\bibnamefont {Tudor}}, \bibinfo
  {author} {\bibfnamefont {A.}~\bibnamefont {Dauphin}}, \bibinfo {author}
  {\bibfnamefont {P.}~\bibnamefont {Massignan}}, \bibinfo {author}
  {\bibfnamefont {L.}~\bibnamefont {Marrucci}},\ and\ \bibinfo {author}
  {\bibfnamefont {F.}~\bibnamefont {Cardano}},\ }\bibfield  {title} {\bibinfo
  {title} {Bloch–{L}andau–{Z}ener dynamics induced by a synthetic field in
  a photonic quantum walk},\ }\href {https://doi.org/10.1063/5.0037327}
  {\bibfield  {journal} {\bibinfo  {journal} {APL Photonics}\ }\textbf
  {\bibinfo {volume} {6}} (\bibinfo {year} {2021})}\BibitemShut {NoStop}%
\bibitem [{\citenamefont {Englebert}\ \emph {et~al.}(2021)\citenamefont
  {Englebert}, \citenamefont {Goldman}, \citenamefont {Erkintalo},
  \citenamefont {Mostaan}, \citenamefont {Gorza}, \citenamefont {Leo},\ and\
  \citenamefont {Fatome}}]{Englebert2021}%
  \BibitemOpen
  \bibfield  {author} {\bibinfo {author} {\bibfnamefont {N.}~\bibnamefont
  {Englebert}}, \bibinfo {author} {\bibfnamefont {N.}~\bibnamefont {Goldman}},
  \bibinfo {author} {\bibfnamefont {M.}~\bibnamefont {Erkintalo}}, \bibinfo
  {author} {\bibfnamefont {N.}~\bibnamefont {Mostaan}}, \bibinfo {author}
  {\bibfnamefont {S.-P.}\ \bibnamefont {Gorza}}, \bibinfo {author}
  {\bibfnamefont {F.}~\bibnamefont {Leo}},\ and\ \bibinfo {author}
  {\bibfnamefont {J.}~\bibnamefont {Fatome}},\ }\bibfield  {title} {\bibinfo
  {title} {Bloch oscillations of driven dissipative solitons in a synthetic
  dimension},\ }\href {https://doi.org/10.48550/arXiv.2112.10756} {\bibfield
  {journal} {\bibinfo  {journal} {arXiv preprint arXiv:2112.10756}\ } (\bibinfo
  {year} {2021})}\BibitemShut {NoStop}%
\bibitem [{\citenamefont {Wang}\ \emph
  {et~al.}(2021{\natexlab{b}})\citenamefont {Wang}, \citenamefont {Dutt},
  \citenamefont {Wojcik},\ and\ \citenamefont {Fan}}]{WangDutt2021}%
  \BibitemOpen
  \bibfield  {author} {\bibinfo {author} {\bibfnamefont {K.}~\bibnamefont
  {Wang}}, \bibinfo {author} {\bibfnamefont {A.}~\bibnamefont {Dutt}}, \bibinfo
  {author} {\bibfnamefont {C.~C.}\ \bibnamefont {Wojcik}},\ and\ \bibinfo
  {author} {\bibfnamefont {S.}~\bibnamefont {Fan}},\ }\bibfield  {title}
  {\bibinfo {title} {Topological complex-energy braiding of non-{H}ermitian
  bands},\ }\href {https://doi.org/10.1038/s41586-021-03848-x} {\bibfield
  {journal} {\bibinfo  {journal} {Nature}\ }\textbf {\bibinfo {volume} {598}},\
  \bibinfo {pages} {59} (\bibinfo {year} {2021}{\natexlab{b}})}\BibitemShut
  {NoStop}%
\bibitem [{\citenamefont {Tusnin}\ \emph {et~al.}(2020)\citenamefont {Tusnin},
  \citenamefont {Tikan},\ and\ \citenamefont {Kippenberg}}]{Tusnin2020}%
  \BibitemOpen
  \bibfield  {author} {\bibinfo {author} {\bibfnamefont {A.~K.}\ \bibnamefont
  {Tusnin}}, \bibinfo {author} {\bibfnamefont {A.~M.}\ \bibnamefont {Tikan}},\
  and\ \bibinfo {author} {\bibfnamefont {T.~J.}\ \bibnamefont {Kippenberg}},\
  }\bibfield  {title} {\bibinfo {title} {Nonlinear states and dynamics in a
  synthetic frequency dimension},\ }\href
  {https://doi.org/10.1103/PhysRevA.102.023518} {\bibfield  {journal} {\bibinfo
   {journal} {Physical Review A}\ }\textbf {\bibinfo {volume} {102}},\ \bibinfo
  {pages} {023518} (\bibinfo {year} {2020})}\BibitemShut {NoStop}%
\bibitem [{not()}]{noteforTBCs}%
  \BibitemOpen
  \href@noop {} {}\bibinfo {note} {The twisted boundary condition is present
  since, in principle, a mode can explore the entire 2D space with only
  nearest-neighbor hoppings. This effectively produces a lattice on the surface
  of a cylinder. See Supplementary materials for details.}\BibitemShut {Stop}%
\bibitem [{\citenamefont {Haldane}(1988)}]{Haldane1988}%
  \BibitemOpen
  \bibfield  {author} {\bibinfo {author} {\bibfnamefont {F.~D.~M.}\
  \bibnamefont {Haldane}},\ }\bibfield  {title} {\bibinfo {title} {Model for a
  quantum {H}all effect without {L}andau levels: Condensed-matter realization
  of the ``{P}arity {A}nomaly"},\ }\href
  {https://doi.org/10.1103/physrevlett.61.2015} {\bibfield  {journal} {\bibinfo
   {journal} {Physical Review Letters}\ }\textbf {\bibinfo {volume} {61}},\
  \bibinfo {pages} {2015} (\bibinfo {year} {1988})}\BibitemShut {NoStop}%
\bibitem [{\citenamefont {Mak}\ \emph {et~al.}(2014)\citenamefont {Mak},
  \citenamefont {McGill}, \citenamefont {Park},\ and\ \citenamefont
  {McEuen}}]{Mak2014}%
  \BibitemOpen
  \bibfield  {author} {\bibinfo {author} {\bibfnamefont {K.~F.}\ \bibnamefont
  {Mak}}, \bibinfo {author} {\bibfnamefont {K.~L.}\ \bibnamefont {McGill}},
  \bibinfo {author} {\bibfnamefont {J.}~\bibnamefont {Park}},\ and\ \bibinfo
  {author} {\bibfnamefont {P.~L.}\ \bibnamefont {McEuen}},\ }\bibfield  {title}
  {\bibinfo {title} {The valley {H}all effect in {MoS}$_2$ transistors},\
  }\href {https://doi.org/10.1126/science.1250140} {\bibfield  {journal}
  {\bibinfo  {journal} {Science}\ }\textbf {\bibinfo {volume} {344}},\ \bibinfo
  {pages} {1489} (\bibinfo {year} {2014})}\BibitemShut {NoStop}%
\bibitem [{\citenamefont {Jim{\'{e}}nez-Gal{\'{a}}n}\ \emph
  {et~al.}(2020)\citenamefont {Jim{\'{e}}nez-Gal{\'{a}}n}, \citenamefont
  {Silva}, \citenamefont {Smirnova},\ and\ \citenamefont
  {Ivanov}}]{JimnezGaln2020}%
  \BibitemOpen
  \bibfield  {author} {\bibinfo {author} {\bibfnamefont {{\'{A}}.}~\bibnamefont
  {Jim{\'{e}}nez-Gal{\'{a}}n}}, \bibinfo {author} {\bibfnamefont {R.~E.~F.}\
  \bibnamefont {Silva}}, \bibinfo {author} {\bibfnamefont {O.}~\bibnamefont
  {Smirnova}},\ and\ \bibinfo {author} {\bibfnamefont {M.}~\bibnamefont
  {Ivanov}},\ }\bibfield  {title} {\bibinfo {title} {Lightwave control of
  topological properties in 2d materials for sub-cycle and non-resonant valley
  manipulation},\ }\href {https://doi.org/10.1038/s41566-020-00717-3}
  {\bibfield  {journal} {\bibinfo  {journal} {Nature Photonics}\ }\textbf
  {\bibinfo {volume} {14}},\ \bibinfo {pages} {728} (\bibinfo {year}
  {2020})}\BibitemShut {NoStop}%
\bibitem [{\citenamefont {Bentsen}\ \emph {et~al.}(2019)\citenamefont
  {Bentsen}, \citenamefont {Hashizume}, \citenamefont {Buyskikh}, \citenamefont
  {Davis}, \citenamefont {Daley}, \citenamefont {Gubser},\ and\ \citenamefont
  {Schleier-Smith}}]{Bentsen2019}%
  \BibitemOpen
  \bibfield  {author} {\bibinfo {author} {\bibfnamefont {G.}~\bibnamefont
  {Bentsen}}, \bibinfo {author} {\bibfnamefont {T.}~\bibnamefont {Hashizume}},
  \bibinfo {author} {\bibfnamefont {A.~S.}\ \bibnamefont {Buyskikh}}, \bibinfo
  {author} {\bibfnamefont {E.~J.}\ \bibnamefont {Davis}}, \bibinfo {author}
  {\bibfnamefont {A.~J.}\ \bibnamefont {Daley}}, \bibinfo {author}
  {\bibfnamefont {S.~S.}\ \bibnamefont {Gubser}},\ and\ \bibinfo {author}
  {\bibfnamefont {M.}~\bibnamefont {Schleier-Smith}},\ }\bibfield  {title}
  {\bibinfo {title} {Treelike interactions and fast scrambling with cold
  atoms},\ }\href {https://doi.org/10.1103/physrevlett.123.130601} {\bibfield
  {journal} {\bibinfo  {journal} {Physical Review Letters}\ }\textbf {\bibinfo
  {volume} {123}} (\bibinfo {year} {2019})}\BibitemShut {NoStop}%
\bibitem [{\citenamefont {Battiston}\ \emph {et~al.}(2021)\citenamefont
  {Battiston}, \citenamefont {Amico}, \citenamefont {Barrat}, \citenamefont
  {Bianconi}, \citenamefont {de~Arruda}, \citenamefont {Franceschiello},
  \citenamefont {Iacopini}, \citenamefont {K{\'{e}}fi}, \citenamefont {Latora},
  \citenamefont {Moreno}, \citenamefont {Murray}, \citenamefont {Peixoto},
  \citenamefont {Vaccarino},\ and\ \citenamefont {Petri}}]{Battiston2021}%
  \BibitemOpen
  \bibfield  {author} {\bibinfo {author} {\bibfnamefont {F.}~\bibnamefont
  {Battiston}}, \bibinfo {author} {\bibfnamefont {E.}~\bibnamefont {Amico}},
  \bibinfo {author} {\bibfnamefont {A.}~\bibnamefont {Barrat}}, \bibinfo
  {author} {\bibfnamefont {G.}~\bibnamefont {Bianconi}}, \bibinfo {author}
  {\bibfnamefont {G.~F.}\ \bibnamefont {de~Arruda}}, \bibinfo {author}
  {\bibfnamefont {B.}~\bibnamefont {Franceschiello}}, \bibinfo {author}
  {\bibfnamefont {I.}~\bibnamefont {Iacopini}}, \bibinfo {author}
  {\bibfnamefont {S.}~\bibnamefont {K{\'{e}}fi}}, \bibinfo {author}
  {\bibfnamefont {V.}~\bibnamefont {Latora}}, \bibinfo {author} {\bibfnamefont
  {Y.}~\bibnamefont {Moreno}}, \bibinfo {author} {\bibfnamefont {M.~M.}\
  \bibnamefont {Murray}}, \bibinfo {author} {\bibfnamefont {T.~P.}\
  \bibnamefont {Peixoto}}, \bibinfo {author} {\bibfnamefont {F.}~\bibnamefont
  {Vaccarino}},\ and\ \bibinfo {author} {\bibfnamefont {G.}~\bibnamefont
  {Petri}},\ }\bibfield  {title} {\bibinfo {title} {The physics of higher-order
  interactions in complex systems},\ }\href
  {https://doi.org/10.1038/s41567-021-01371-4} {\bibfield  {journal} {\bibinfo
  {journal} {Nature Physics}\ }\textbf {\bibinfo {volume} {17}},\ \bibinfo
  {pages} {1093} (\bibinfo {year} {2021})}\BibitemShut {NoStop}%
\bibitem [{\citenamefont {Disa}\ \emph {et~al.}(2021)\citenamefont {Disa},
  \citenamefont {Nova},\ and\ \citenamefont {Cavalleri}}]{Disa2021}%
  \BibitemOpen
  \bibfield  {author} {\bibinfo {author} {\bibfnamefont {A.~S.}\ \bibnamefont
  {Disa}}, \bibinfo {author} {\bibfnamefont {T.~F.}\ \bibnamefont {Nova}},\
  and\ \bibinfo {author} {\bibfnamefont {A.}~\bibnamefont {Cavalleri}},\
  }\bibfield  {title} {\bibinfo {title} {Engineering crystal structures with
  light},\ }\href {https://doi.org/10.1038/s41567-021-01366-1} {\bibfield
  {journal} {\bibinfo  {journal} {Nature Physics}\ }\textbf {\bibinfo {volume}
  {17}},\ \bibinfo {pages} {1087} (\bibinfo {year} {2021})}\BibitemShut
  {NoStop}%
\bibitem [{\citenamefont {Kippenberg}\ \emph {et~al.}(2018)\citenamefont
  {Kippenberg}, \citenamefont {Gaeta}, \citenamefont {Lipson},\ and\
  \citenamefont {Gorodetsky}}]{Kippenberg2018}%
  \BibitemOpen
  \bibfield  {author} {\bibinfo {author} {\bibfnamefont {T.~J.}\ \bibnamefont
  {Kippenberg}}, \bibinfo {author} {\bibfnamefont {A.~L.}\ \bibnamefont
  {Gaeta}}, \bibinfo {author} {\bibfnamefont {M.}~\bibnamefont {Lipson}},\ and\
  \bibinfo {author} {\bibfnamefont {M.~L.}\ \bibnamefont {Gorodetsky}},\
  }\bibfield  {title} {\bibinfo {title} {Dissipative {K}err solitons in optical
  microresonators},\ }\href {https://doi.org/10.1126/science.aan8083}
  {\bibfield  {journal} {\bibinfo  {journal} {Science}\ }\textbf {\bibinfo
  {volume} {361}} (\bibinfo {year} {2018})}\BibitemShut {NoStop}%
\bibitem [{\citenamefont {Wright}\ \emph {et~al.}(2017)\citenamefont {Wright},
  \citenamefont {Christodoulides},\ and\ \citenamefont {Wise}}]{Wright2017}%
  \BibitemOpen
  \bibfield  {author} {\bibinfo {author} {\bibfnamefont {L.~G.}\ \bibnamefont
  {Wright}}, \bibinfo {author} {\bibfnamefont {D.~N.}\ \bibnamefont
  {Christodoulides}},\ and\ \bibinfo {author} {\bibfnamefont {F.~W.}\
  \bibnamefont {Wise}},\ }\bibfield  {title} {\bibinfo {title} {Spatiotemporal
  mode-locking in multimode fiber lasers},\ }\href
  {https://doi.org/10.1126/science.aao0831} {\bibfield  {journal} {\bibinfo
  {journal} {Science}\ }\textbf {\bibinfo {volume} {358}},\ \bibinfo {pages}
  {94} (\bibinfo {year} {2017})}\BibitemShut {NoStop}%
\bibitem [{\citenamefont {Karpov}\ \emph {et~al.}(2019)\citenamefont {Karpov},
  \citenamefont {Pfeiffer}, \citenamefont {Guo}, \citenamefont {Weng},
  \citenamefont {Liu},\ and\ \citenamefont {Kippenberg}}]{Karpov2019}%
  \BibitemOpen
  \bibfield  {author} {\bibinfo {author} {\bibfnamefont {M.}~\bibnamefont
  {Karpov}}, \bibinfo {author} {\bibfnamefont {M.~H.~P.}\ \bibnamefont
  {Pfeiffer}}, \bibinfo {author} {\bibfnamefont {H.}~\bibnamefont {Guo}},
  \bibinfo {author} {\bibfnamefont {W.}~\bibnamefont {Weng}}, \bibinfo {author}
  {\bibfnamefont {J.}~\bibnamefont {Liu}},\ and\ \bibinfo {author}
  {\bibfnamefont {T.~J.}\ \bibnamefont {Kippenberg}},\ }\bibfield  {title}
  {\bibinfo {title} {Dynamics of soliton crystals in optical microresonators},\
  }\href {https://doi.org/10.1038/s41567-019-0635-0} {\bibfield  {journal}
  {\bibinfo  {journal} {Nature Physics}\ }\textbf {\bibinfo {volume} {15}},\
  \bibinfo {pages} {1071} (\bibinfo {year} {2019})}\BibitemShut {NoStop}%
\bibitem [{\citenamefont {Haus}(1983)}]{Haus1983}%
  \BibitemOpen
  \bibfield  {author} {\bibinfo {author} {\bibfnamefont {H.~A.}\ \bibnamefont
  {Haus}},\ }\href@noop {} {\emph {\bibinfo {title} {Waves and Fields in
  Optoelectronics}}},\ Prentice-Hall series in solid state physical
  electronics\ (\bibinfo  {publisher} {Prentice Hall},\ \bibinfo {address} {Old
  Tappan, NJ},\ \bibinfo {year} {1983})\BibitemShut {NoStop}%
\bibitem [{\citenamefont {Groth}\ \emph {et~al.}(2014)\citenamefont {Groth},
  \citenamefont {Wimmer}, \citenamefont {Akhmerov},\ and\ \citenamefont
  {Waintal}}]{Groth2014}%
  \BibitemOpen
  \bibfield  {author} {\bibinfo {author} {\bibfnamefont {C.~W.}\ \bibnamefont
  {Groth}}, \bibinfo {author} {\bibfnamefont {M.}~\bibnamefont {Wimmer}},
  \bibinfo {author} {\bibfnamefont {A.~R.}\ \bibnamefont {Akhmerov}},\ and\
  \bibinfo {author} {\bibfnamefont {X.}~\bibnamefont {Waintal}},\ }\bibfield
  {title} {\bibinfo {title} {Kwant: a software package for quantum transport},\
  }\href {http://dx.doi.org/10.1088/1367-2630/16/6/063065} {\bibfield
  {journal} {\bibinfo  {journal} {New Journal of Physics}\ }\textbf {\bibinfo
  {volume} {16}},\ \bibinfo {pages} {063065} (\bibinfo {year}
  {2014})}\BibitemShut {NoStop}%
\end{thebibliography}%

\appendix

\newpage

\begin{center}
\textbf{\large Supplementary Materials: Programmable large-scale simulation of bosonic transport in optical synthetic frequency lattices}
\end{center}

\setcounter{equation}{0}
\setcounter{figure}{0}
\setcounter{table}{0}
\setcounter{page}{1}
\makeatletter
\renewcommand{\theequation}{S\arabic{equation}}
\renewcommand{\thefigure}{S\arabic{figure}}

\subsection{Experimental setup}

\begin{figure}[h]
    \centering
    \includegraphics[width=.9\linewidth]{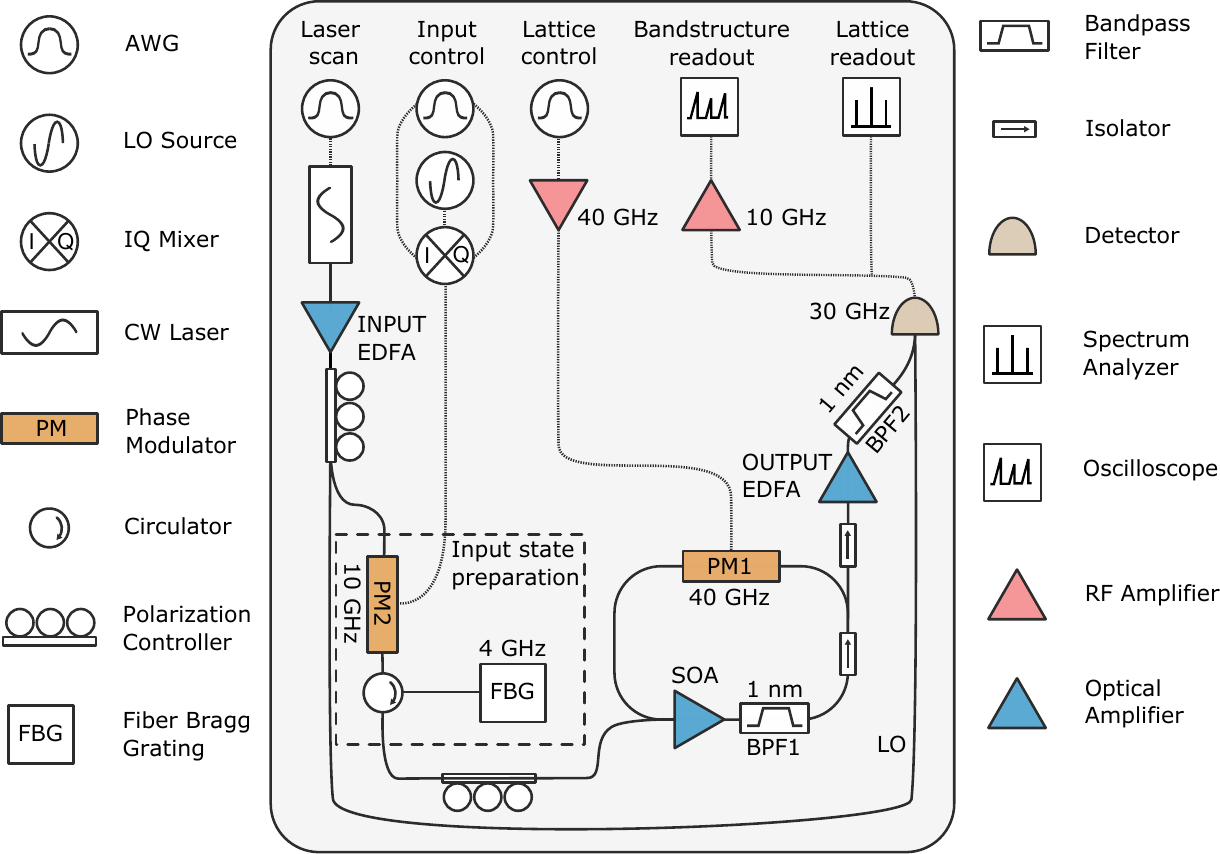}
    \caption{Schematic of the experimental setup. }
    \label{fig:exp_setup}
\end{figure}

\begin{figure}[h]
    \centering
    \includegraphics[width=.9\linewidth]{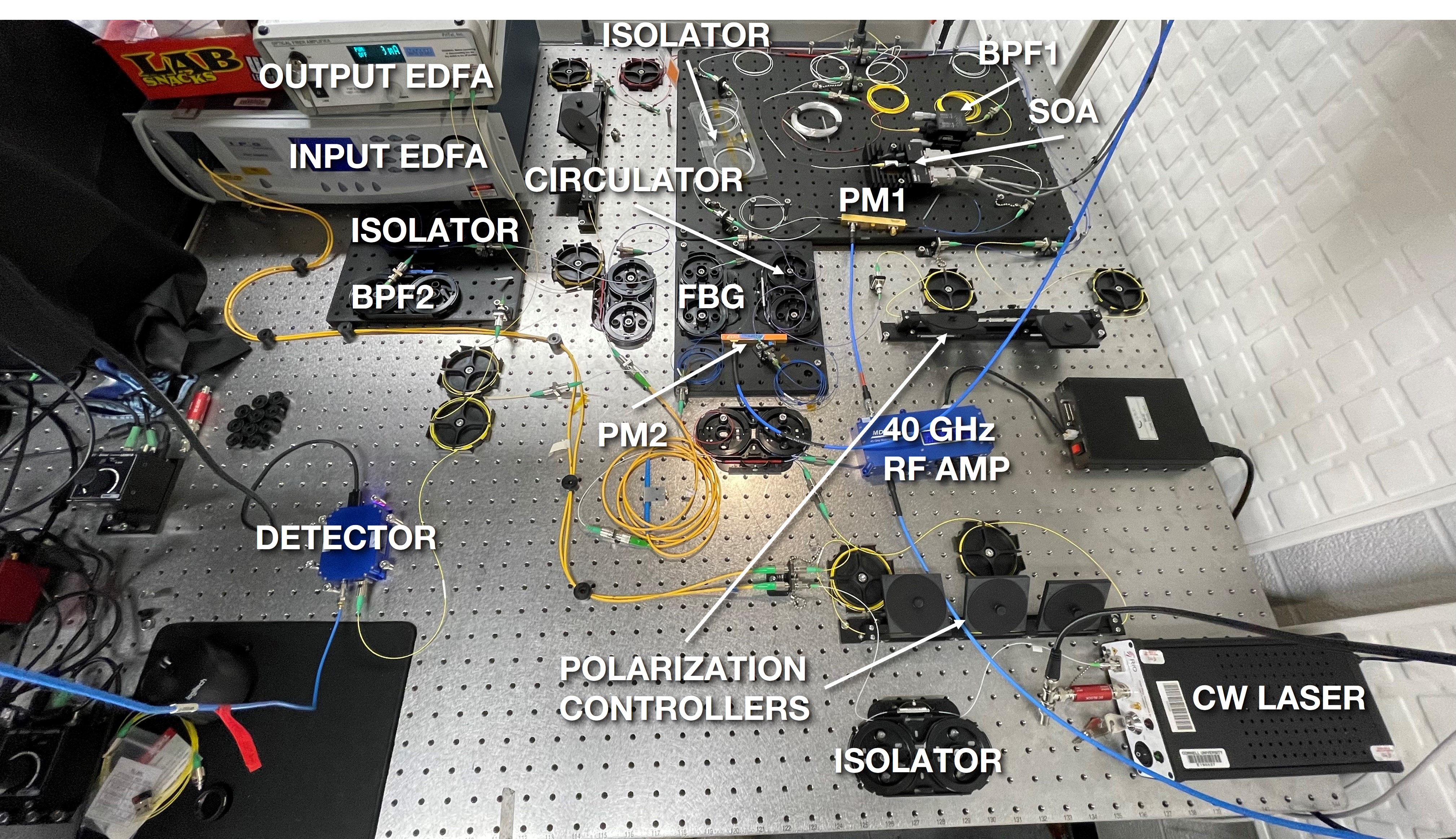}
    \caption{Photograph of the experimental setup.}
    \label{fig:exp_setup_pic}
\end{figure}

The experimental setup, schematically shown in Fig.~\ref{fig:exp_setup}, consists of a long fiber ring with a free spectral range (FSR) $\Omega = 1.226045$ MHz, modulated with a 40-GHz phase modulator (LN27S-FC from Thorlabs) to produce the lattices in the main text and below. The losses in the cavity are compensated by an semiconductor optical amplifier (SOA) from Thorlabs (SOA1117S). A 1 nm optical filter (OZ Optics BTF-11-11-1525/1565-9) is placed in the cavity to reduce ASE generated by the SOA, and an isolator (Thorlabs IO-H-1550APC) is placed to reduce parasitic resonances particularly at low cavity power. 

In addition to the cavity, we implement arbitrary input states by a filtering modulations of an injection. The injection laser (RIO3335-3-00-1-BZ7), centered at 1550.57 nm, is first amplified before being split with a 50:50 coupler. One arm serves as the local oscillator (LO) for heterodyne detection, and the other arm is sent through a phase modulator (EOSpace) to prepare the input state. The polarization is adjusted so that it is aligned with the crystal axis of the phase modulator. After passing through the modulator, the input state is encoded in sidebands centered 12 GHz away from the injection. A fiber Bragg grating (FBG) mirror reflects a 4 GHz portion of the spectrum centered 12 GHz away from the LO, while the rest, containing the other sideband and the LO frequency, is dumped (see Fig.~\ref{fig:figure3}A). We place a circulator in between the phase modulator and the FBG to collect the reflected light. While this prescription results in 20 dB of insertion loss, we find 30 dB of isolation between the filtered sideband and the LO resulting in a clean single-side band modulation. The modulation of both the input state phase modulator and the cavity phase modulator are controlled by a single AWG (Keysight M8195A). The polarization of the prepared input state programmed by a dedicated AWG is set to align with the crystal axis of the intra-cavity EOM. 

Finally, the output of the cavity is combined with the LO before being detected by a 30-GHz photodetector (Optilab PD-30-M-K-DC), resulting in a 12-GHz heterodyne detection of the output. One component of resultant RF signal is amplified before sent to an oscilloscope (Tektronix DSA72004) for band structure spectroscopy in the time-domain.  The other component of the RF signal is sent to the spectrum analyzer (Tektronix RSA5126A) to perform direct readout of the lattice over 26 GHz. For bandstructure measurements, the heterodyne arm is turned off. The spectrum measured in Fig.~\ref{fig:figure5}B was measured with an optical spectrum analyzer (Ando AQ6317). 

\begin{figure}[h]
    \centering
    \includegraphics[width=.75\textwidth]{ 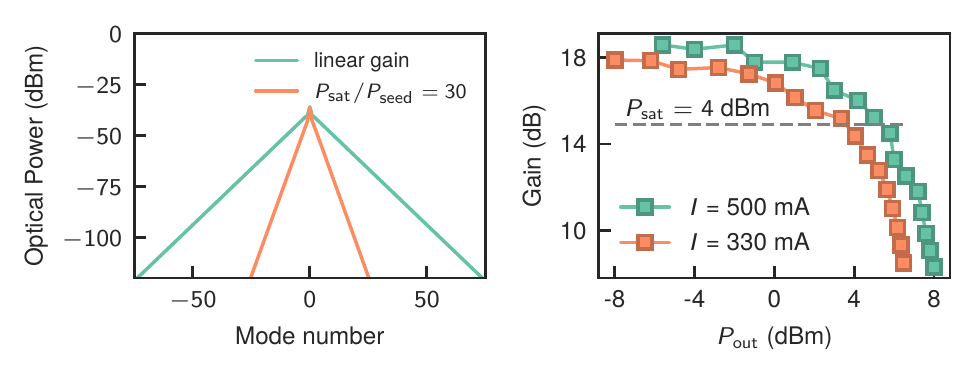}
    \caption{Modeling and characterization of gain saturation. Left: simulations of cavity dynamics for linear gain (green curve) and saturating gain (orange curve). Right: Experimental characterization of semiconductor amplifier at various levels of pump current. }
    \label{fig:soa_characterization}
\end{figure}

\subsection{Setup characterization}

Here we outline characterizations of the cavity, particularly characterizations of the cavity gain and losses, coherence time, and the free spectral range (FSR). 

The intracavity SOA compensates for roundtrip losses in the cavity. However, the SOA contributes significantly to noise by the unwanted production of ASE. In order to reduce ASE, we placed a filter in the cavity, and minimized roundtrip losses to reduce the operating point of the SOA. The cavity losses were reduced to 5 dB, 4 dB of which originate from the insertion loss of the cavity EOM. In addition to limiting the noise, reducing the operating point of the SOA has the added benefit of reducing the contribution of the ASE to the gain saturation. The left panel of Fig.~\ref{fig:soa_characterization} shows the effect of the gain saturation on the dynamics of a nearest-neighbor optical lattice. For a large ratio of input power to saturating power, the transport of the optical power along the 1D chain is limited. From simulations, we find that a ratio of $P_{\text{seed}}/P_{\text{sat}} < 1/100$ is enough to reach thousands of modes above the noise floor of our setup. The right side of Fig.~\ref{fig:soa_characterization} shows the experimental measured value of gain saturation at about 4 dBm, thus for all of the measurements in the main text, we had an input power of below -20 dBm going into the cavity. 

\begin{figure}[h]
    \centering
    \includegraphics{ 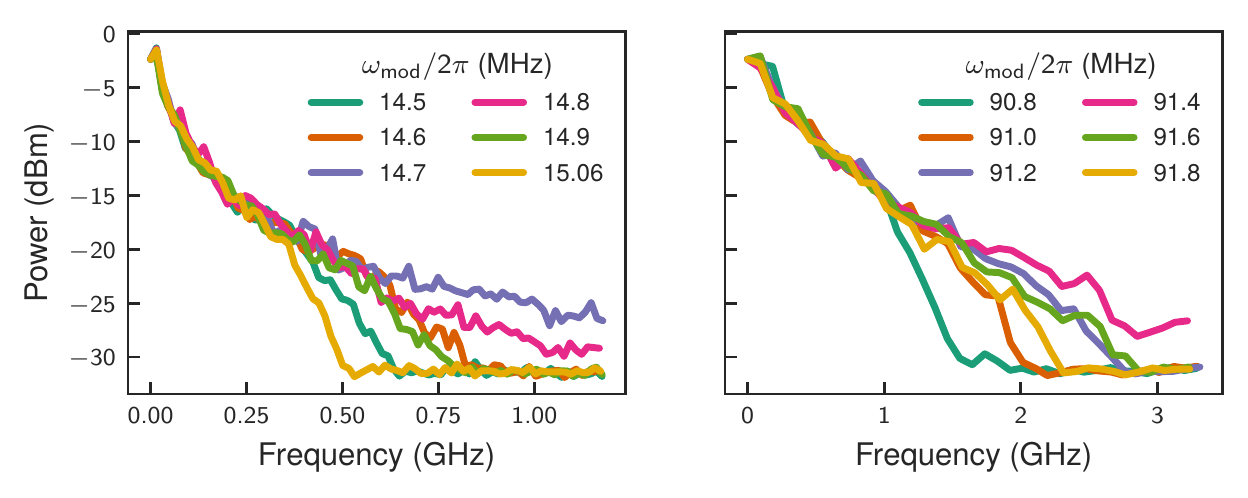}
    \caption{Measurement of the cavity resonance by maximizing transport response at lasing threshold. The cavity was modulated around a window of $\omega_{\mathrm{mod}} \approx 12 \times \Omega$ (Left), and $\omega_{\mathrm{mod}} \approx 75 \times \Omega$ (Right). }
    \label{fig:1d_resonance}
\end{figure}

\begin{figure}[h]
    \centering
    \includegraphics{ 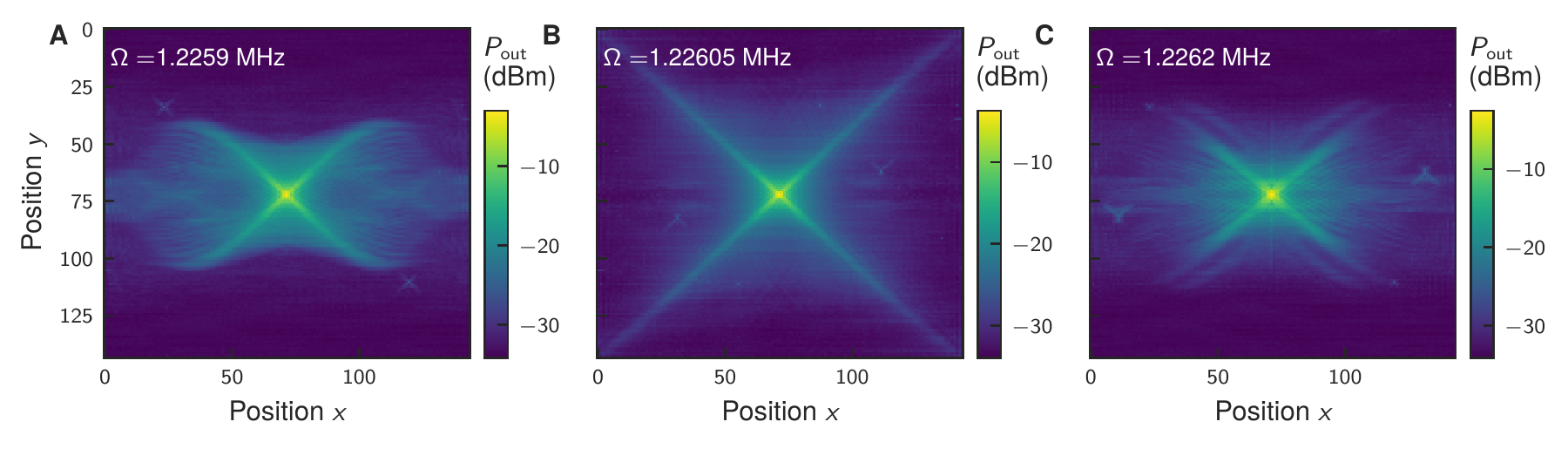}
    \caption{Response to single site injection in a 2D square lattice with 20,000 sites as a function of detuning between the modulation frequency and the cavity free spectral range. By optimizing the detuning for transport, the cavity FSR was measured to be 1.22605 MHz. Panels A and B show the presence of Bloch oscillations (see Supplementary Section \ref{sec:supp_results})   }
    \label{fig:2d_resonance}
\end{figure}

Knowledge of the cavity FSR is needed to $\sim$ 10 Hz to in order sustain the coherence over many roundtrips needed to instantiate large lattices. This was measured via different methods, differing in course-grained vs fine-grained characterization. At the first step, we measured the FSR by exciting the cavity far above the lasing threshold and measuring the mode excitations. This procedure gives us the FSR to within 10 kHz. Next, we placed the cavity just below the lasing threshold and performed spectral measurements due to a single-site injection, and maximized the transport observed over the modes as we varied the modulation frequency. Shown in Fig.~\ref{fig:1d_resonance} are the responses in a 1D chain as a function of modulation frequency at cavity threshold. To further increase the sensitivity of the measurement, the same procedure was applied at larger integer multiple harmonics. These are plotted in Fig.~\ref{fig:1d_resonance}B.

To measure the FSR of our cavity down to 10 Hz, we modulated the EOM with multiple tones, which increased the sensitivity due to interferences between different paths taken over multiple roundtrips of the cavity. For example, a modulation consisting of two tones at $\Omega$ and $10\times \Omega$ will maximize transport if 10 hoppings along the nearest-neighbor modulation produces is coherent with a single hop along $10\times \Omega$. Figure~\ref{fig:2d_resonance} shows the transport response for a $144 \times 144$ 2D lattice instantiated by two tone modulations at frequencies $\omega_{\mathrm{mod}}$ and $144\times \omega_{\mathrm{mod}}$. When detuned from the FSR (left and right panels of Fig.~\ref{fig:2d_resonance}), we see the injected light reaching out then oscillating back in. Closer to resonance (middle panel of Fig.~\ref{fig:2d_resonance}), the light propagates symmetrically outward. This effect is related to Bloch oscillations (see Section~\ref{sec:supp_results}), expected to occur in modulated cavity systems where the modulation is detuned  \cite{Yuan2016}. Suppressing these oscillations allowed us to find the FSR to 5 decimal places, down to 10 Hz.

\begin{figure}[h]
    \centering
    \includegraphics[width=\linewidth]{ 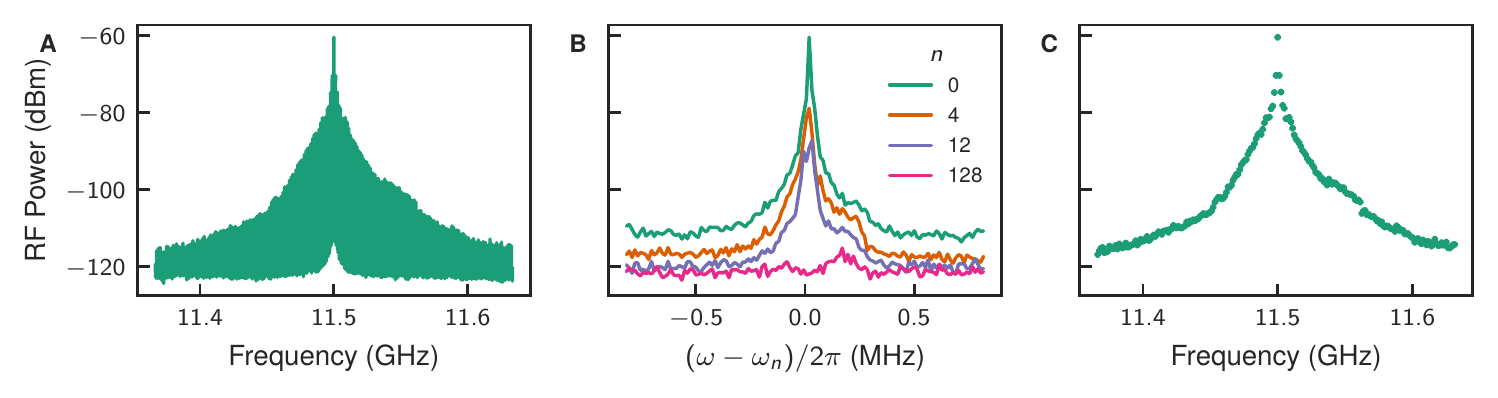}
    \caption{Processing raw spectrum data. (A) RF spectral cavity response due to a single site injection in a 1D nearest-neighbor chain. (B) closeup of mode excitation for various lattice sites overlayed on the same scale. $n$ is the distance away from the injection measured in number of lattice sites. (C) Constructed lattice plot from peaks of lattice site excitation.  }
    \label{fig:peak_picking}
\end{figure}

\subsection{Setup modeling }
\label{sec:setup_modeling}

Here we outline a simple model for the cavity dynamics. We consider the complex-valued electric field $A(x,t)$ inside the ring cavity cavity such that $A(x,t) = A(x+L,t)$ where $L$ is the roundtrip length of the cavity.  In the absence of dispersion, we can expand this electric field in terms of resonant modes as $A(x,t) = \sum_m A(x)a_m(t)e^{im\Omega t}$~\cite{Haus1983} due to this periodic boundary condition. Here $\Omega$ is the mode spacing, related to the roundtrip length as $\Omega = c/L$. In the frequency domain, the resonant modes $a_m$ are then coupled via an EOM to realize a variety of tight-binding graphs. In the time-domain, the action of the phase modulator over one roundtrip is simply $a(t+\tau) = e^{i\phi(t)}a(t)$ where the cavity modulation $\phi(t)$ is proportional to the RF signal $v(t)$ driving the phase modulator and is periodic over the roundtrip time $\tau$. In the basis of the frequency modes, this action is described by
\begin{equation}
    U = e^{iJ\tau} = F^{\dagger}e^{i V } F
\end{equation}
where $V$ is a diagonal operator encoding the modulation $\phi(t)$ along the diagonal, and $F$ is the discrete Fourier transform $F_{jk} = (e^{-i2\pi/N})^{jk}$, where $N$ is the number of modes. The equation above sets constraints on the time evolution operator $U$, namely that it is a unitary Toeplitz operator, i.e. a unitary matrix with constant diagonals. 

For small values of the matrix $J$, the full coupled mode equations are

\begin{equation}
\dot{a_m}(t) = \left(im\Omega - \frac{g}{1 + \sum_n \vert a_n \vert^2/P_{\mathrm{sat}}} - \ell\right) a_m(t) - i\sum_n J_{n-m} a_n(t) + a_{\mathrm{in}}(t)e^{i\Delta t},
\end{equation}
where $g$ is the small-signal gain, $\ell$ is the roundtrip loss, $a_{\mathrm{in}}$ is a complex frequency dependant amplitude encoding the input state at frequency $\omega_{\mathrm{in}}$, and $P_{\text{sat}}$ is the nonlinear saturating gain. The $J_k$ describe the coupling terms from the phase modulator, given by $J_{k} = (2\pi)^{-1} \int_0^{1/\Omega}dt e^{i\phi(t)} e^{i k \Omega t}$. To model the effect of dispersion we would include a term proportional to $m^2$ in the first term, but we have not done so because in our experiments the dispersion was sufficiently small that it could be neglected.

This equation is simplified by moving to the rotating frame of each mode $a_m(t) \rightarrow a_m(t)e^{i m \Omega t}$ and operating well below the saturation power. In this form, the equations reduce to
\begin{equation}
\dot{a_m}(t) = -\gamma a_m(t) - i\sum_n J_{n-m} a_n(t) + a_{\mathrm{in}}(t)e^{i\Delta t},
\end{equation}
where $\gamma = \ell - g$ is the gain-loss balance, and $\Delta$ is the detuning from the cavity modes. Collecting the mode amplitudes in a column vector $\vert a (t) \rangle = (a_0(t), a_1(t) ...)^T$ with the basis $\{\vert m \rangle\}$ indexing the cavity modes, we obtain
\begin{equation}
    i \langle m \vert \left( \frac{\partial}{\partial t} - i \gamma + J  \right)\vert a (t) \rangle = \langle m \vert a_{\mathrm{in}} \rangle e^{i\Delta t}.
    \label{eq:eoms_matrixform}
\end{equation}
Neglecting the loss in the system, this equation describes the Schrodinger equation derived from the Hamiltonian in Eq. \eqref{eq:hamiltonian} in the main text, with $H = J$. The implicit assumption is that since the Hamiltonian is sufficiently linear, the many-body states are described with just a single-excitation picture, i.e. with first-quantization.  

Equation \eqref{eq:eoms_matrixform} has a simple steady state solution in the continuous limit for a single site injection, i.e. $\vert a_{\mathrm{in}} \rangle = (0,0,...0,1,0,...,0,0)^T$. Substituting the basis vectors $\{\langle m \vert \}$ indexing the mode numbers with a continuous parameter $x$, we can integrate Eq. \eqref{eq:eoms_matrixform} with Fourier transforms to obtain
\begin{equation}
    a(x,t\rightarrow \infty) = i \sqrt{P_{\mathrm{in}}}\int \frac{dk  \ e^{ikx}} { 
    \phi_k- i \gamma }
    \label{eq:sol_gen}
\end{equation}
where $\phi_k$ is the spectrum of $J$. For example, a 1D nearest neighbor coupling with strength $J_0$, where $\phi_k = k^2$, Eq. \eqref{eq:sol_gen} has the steady-state solution 
\begin{equation}
    a(x) = \pi \sqrt{P_{\mathrm{in}}} \frac{e^{\sqrt{2\gamma/J_0}\vert x\vert }}{2\gamma/J_0}
    \label{eq:sol}
\end{equation}
where $x$ is the continuous variable indexing the mode number. Clearly, by taking $\gamma\rightarrow 0$, i.e. operating at threshold, the number of occupied modes reaches infinity. Of course, the solution in Eq. \eqref{eq:sol} neglects the effects of dispersion and more importantly, gain saturation. From simulations, we found that dispersion would begin limiting our lattice sizes around a bandwidth of $\sim$ 100 GHz, but that the gain saturation limits the size much earlier, effectively increasing the loss as discussed above.

Simulations used to compare with experimental results were performed using a driven-dissipative tight-binding Hamiltonian with the true underlying geometry. That is, if we are comparing a 2D lattice in our experimental system with one from simulations, the Hamiltonian implementing the lattice in the computer simulations is a true 2D lattice without the twisted boundary conditions arising from the translational invariant coupling nature of the phase modulators. This is to directly compare the performance of the simulator vs computer simulations of these systems. In addition, the effects of the gain nonlinearity were neglected, other than to induce an effective decrease in the gain of the amplifier (so that the modes don't populate out to infinity). In short, with the exception of Fig.~\ref{fig:figure5}, the simulations performed in this work were modeled with Eq.~\eqref{eq:eoms_matrixform}, with the coupling matrix $J$ replaced with the true underlying tight-binding Hamiltonian. Since the lattice in Fig.~\ref{fig:figure5} is a 1D lattice with non-local connections, and is not emulating some higher-dimensional planar graph, this was simulated faithfully. 

\subsection{Real-space-occupation \& band-structure measurements}
\label{sec:supp_bs}

Here we provide the methods used in measurements of the steady state tomography in various lattices. The output of the cavity is sent to an RF detector, whose current measures the optical power. This RF signal is then sent to an RF spectrum analyzer which integrates the signal over 100s of round trips. The spectrum analyzer measures RF power, related to the optical power via $P_{\mathrm{RF}} = \eta^2 P_{\mathrm{opt}}^2 \times 50 \Omega$, where $\eta$ is the responsitivity of the photodetector (measured to be about $0.77$ A/W). All spectral plots are plotted in terms of optical power leaving the cavity before the EDFA, with the exception of Fig.~\ref{fig:peak_picking}, which plots the raw RF power spectrum. 

To map the power spectrum to the occupation at a particular lattice site, the peak of the power around a neighborhood of the expected frequency of the cavity mode was picked. Figure~\ref{fig:peak_picking}A shows the raw heterodyne response data for a 1D lattice due to a single-site injection. Figure~\ref{fig:peak_picking}B shows several cavity modes imposed on the same axis for a neighborhood of 1 MHz just below threshold. Here, the linewidth of the cavity is about 0.5 MHz. From these windows, the peaks were picked to construct the final lattice response, displayed in the right pane of Fig.~\ref{fig:peak_picking}. 

To generate lattice plots in two dimensions and higher, we employed the above prescription to retrieve the peaks from a 1D chain, then reshaped the data appropriately depending on the type of lattice we were realizing. For example, if we measure the response in a 2D square lattice, where we modulate the cavity at $\Omega$ and $L\Omega$, we pick $N = L \times L$ peaks of the spectrum, and wrap them in a $L\times L$ matrix. 

\begin{figure}
    \centering
    \includegraphics[width=.9\linewidth]{ 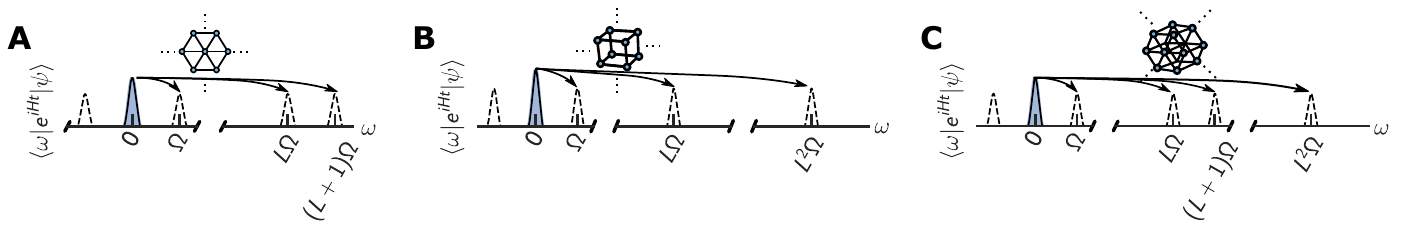}
    \caption{ Modulation schemes for the lattices presented in Fig.~\ref{fig:figure2} of the main text: (A) 2D triangular lattice, (B) 3D square lattice, (C) 3D triangular lattice. The modulation scheme for a 2D square lattice is shown in Fig.~\ref{fig:figure1}. }
    \label{fig:modulation_schemes}
\end{figure}

\begin{figure}
    \centering
    \includegraphics{ 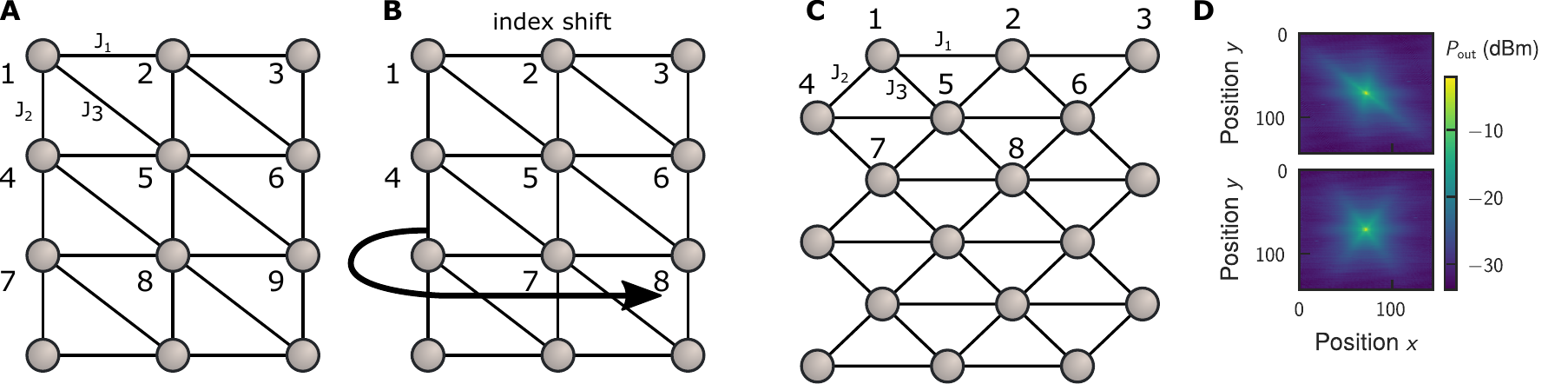}
    \caption{Mapping the frequency spectrum to a 2D triangular lattice geometry. First, the frequency modes are mapped to points in (A), similar to a 2D square lattice, though with an extra hopping. Beginning with every other row, the indices are shifted over by one in order to preserve the local geometry of the triangular lattice (B). Finally, every other row is shifted by half a "pixel" in all spatial representations (C). (D) Compares the unshifted lattice (top) with the shifted (bottom).  }
    \label{fig:Triangular_index_shift}
\end{figure}

For a triangular lattice, we modulate the intra-cavity EOM at three frequencies $\{ \Omega, L\Omega, (L+1) \Omega \}$ (Fig.~\ref{fig:modulation_schemes}A). Following the methods employed in the 2D square lattice geometry, this produces a lattice depicted in Fig.~\ref{fig:Triangular_index_shift}A -- a 2D square lattice with an extra diagonal connections.  While the connectivity of this graph is exact with the connectivity of a triangular lattice, the six fold symmetry of this lattice is not captured when presented in regular euclidean space. Intuitively, this can be seen in Fig.~\ref{fig:Triangular_index_shift}C, where a traversal down along $J_2$ followed by a traversal along $J_3$ should result in an occupation that is directly "below" the original site geometrically. In Fig.~\ref{fig:Triangular_index_shift} however, this procedure results in a shift along the horizontal direction. Thus, beginning with every even row of our lattice, we shift the indices over by one as depicted in Fig.~\ref{fig:Triangular_index_shift}B, resulting in the expectant traversals. Doing so preserves the local connectivity of the triangular lattice without physically altering the connectivity. Finally, when plotting these lattices on a 2D plane, we plot the values of the modes on a planar lattices that has the geometry depicted in Fig.~\ref{fig:Triangular_index_shift}C. The difference in these aesthetic changes are apparent in Fig.~\ref{fig:Triangular_index_shift}D. 

\begin{figure}
    \centering
    \includegraphics{ 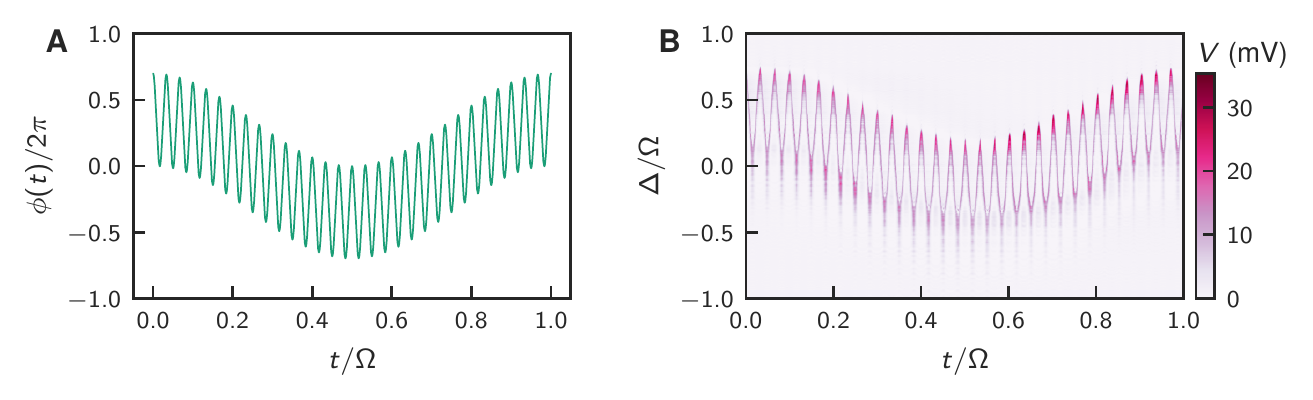}
    \caption{Bandstructure measurement of a 30 $\times$ 30 2D square lattice using methods introduced in \cite{Dutt2019}. (A) Modulation signal sent to the cavity EOM. (B) Experimental measurement of time-domain response of the cavity broken up in chunks of the roundtrip time $\tau = 1/\Omega$. The vertical axis on the experimental plots is the detuning $\Delta$ of the injection away from the cavity mode, normalized by the cavity spacing $\Omega$. }
    \label{fig:bs_measurement}
\end{figure}

\begin{figure}
    \centering
    \includegraphics{ 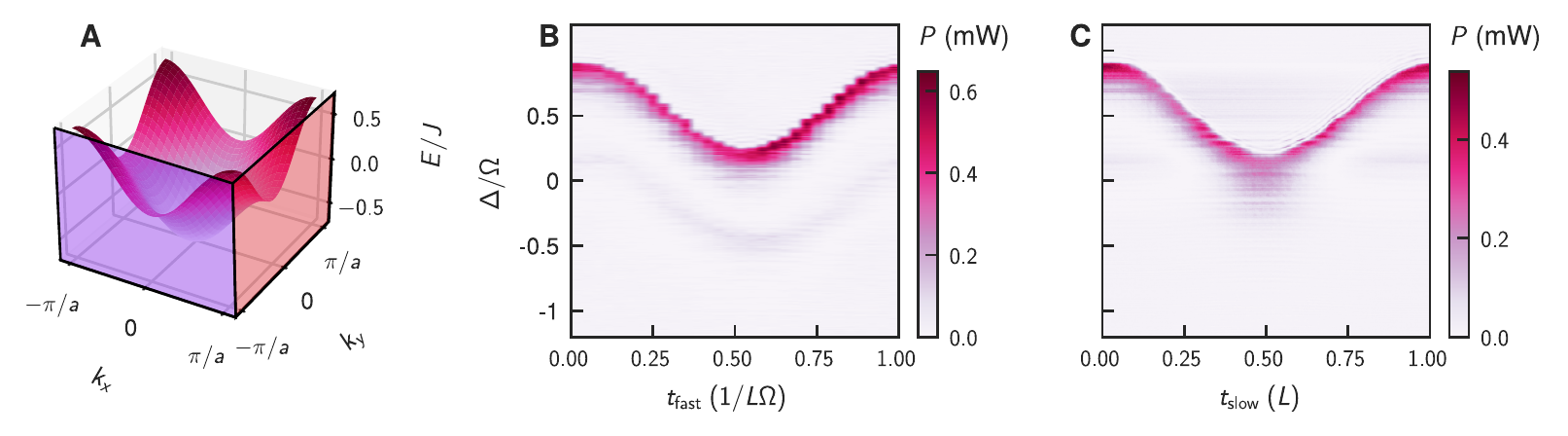}
    \caption{Decomposition of band structure of a 2D square lattice along edges of the Brillouin zone. (A) Analytic band structure for a 2D square lattice, with edges of the Brillouin zone highlighted with purple and red planes. By slicing the measured band structures into $L$ chunks along the fast time axis each of length $t_{\mathrm{fast}} = \Omega/L$, we can reconstruct the full 2D band structure. 
    Slices of the reconstructed 2D band structure corresponding to $k_x = \pi/a$ (B), and $k_y = \pi/a$ (C). }
    \label{fig:bs_decomp}
\end{figure}

\begin{figure}
    \centering
    \includegraphics[width=.7\linewidth]{ 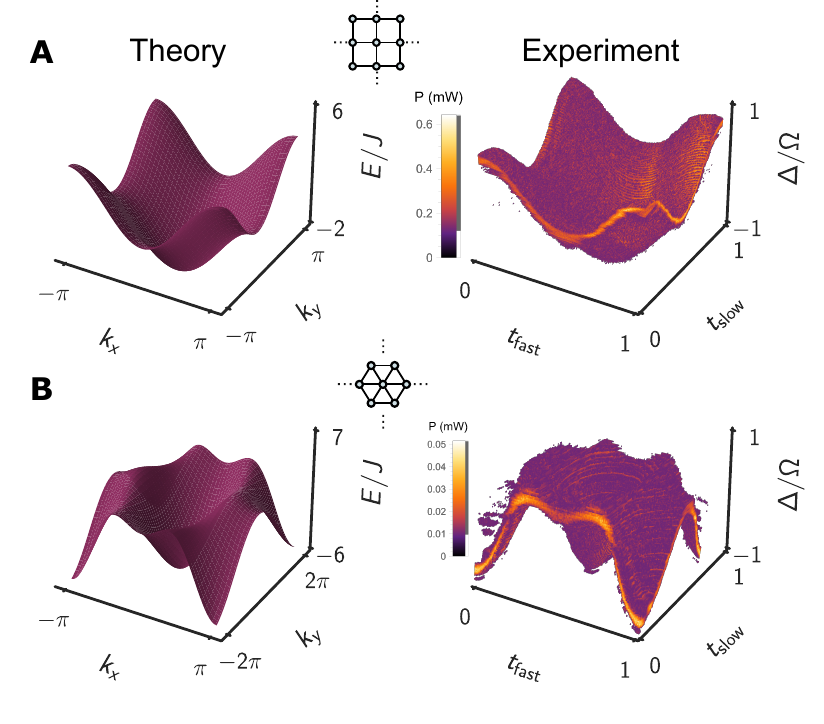}
    \caption{Full reconstruction of band structure measurements for modulations instantiating lattices with $> 10,000$ lattice sites. (A) band structure for a 2D square lattice comparing theory with experiment. (B) band structure for a 2D triangular lattice comparing theory with experiment. The experimental data is in effect a function of three variables. A opacity cutoff has been applied at low signal for clarity (shown in the grey bar adjacent to the colorbar).}
    \label{fig:2d_bandstructure}
\end{figure}

To measure band structure, the output of the cavity is amplified and filtered with a 0.1-1 GHz-bandpass filter. The amplifier increases the signal to noise ratio needed for the single shot readout for the band structure measurements. Single shot read-out is needed due to the phase walk-off between the injection and the cavity.  For details of the band structure measurement, we refer to~\cite{Dutt2019}, but we briefly summarize the procedure here, and outline our extensions for measurements of band structures in 2D and higher. 

Figure~\ref{fig:bs_measurement} shows the modulation signal sent to the cavity to realize a $30 \times 30$ 2D square lattice, and the measured band structure spectroscopy. The plot on the right is produced by taking a linear scan of the injection signal over a cavity mode, and measuring the response. The raw measurement is a 1D time series, which is then divided up into chunks set by the cavity roundtrip time $\tau = 1/\Omega$. The vertical axis is set by the scanning speed, here normalized by the cavity mode spacing $\Omega$. Here, we scanned over one mode in 1 ms. These two time scales were observed to be well separated enough to allow the laser to equilibrium with the continuously changing scanning frequency. The widths of the band structures are proportional to $\Omega V/V_{\pi}$, here $V$ is the modulation amplitude, and $V_{\pi}$ is the pi-voltage of the phase modulator. For all band structure measurements, we drove the EOM very close to the pi-voltage in order to get wide band structures. 

A real 2D square lattice has a band structure of $E (k) = -2 J \cos(k_x a) - 2 J \cos(k_y a)$. Here, we have a single dimension. To construct the full 2D band structure from this signal, we further separate the time-domain response signal into two further time scales by decomposing the horizontal axis of Fig.~\ref{fig:bs_measurement}B into $L$ chunks, so that each chunk is of length $t_{\mathrm{fast}} = 1/L\Omega$, where $L$ is the secondary long range coupling used to instantiate a 2D square lattice. The left plot of Fig.~\ref{fig:bs_decomp} shows the first chunk of the data plotted in Fig.~\ref{fig:bs_measurement}. The secondary time scale is synthetically formed by looking at points separated by $1/L\Omega$. In other words, if we reconstruct the full 2D band structure by appending chunks of length $t_{\mathrm{fast}}$, $t_{\mathrm{slow}}$ is the orthogonal direction pointing along the different chunks. These two timescales $t_{\mathrm{fast}}, t_{\mathrm{slow}}$ map to the two independent momenta $k_x$, and $k_y$, when the time scales are well separated enough, as is the case for $L \gg 1$. Figures~\ref{fig:bs_decomp}B \& C show the band structures along $t_{\mathrm{fast}}$ and $t_{\mathrm{slow}}$, corresponding to borders of the Brillouin zone with $k_x = \pi/a$ and $k_y = \pi/a$, respectively.

Finally, Fig.~\ref{fig:2d_bandstructure} shows the full reconstructed 2D band structure for a square and triangular lattice, along with the analytic band structures. For clarify, an opacity filter was applied to the experimental data for the 3D plots in Fig.~\ref{fig:2d_bandstructure} such that only points greater than a certain power level are plotted. The band structures shown in the main text are slices of the full reconstructed band structure along high symmetry points. The theory curves were computed from the analytically solvable band structures, namely $E_{\mathrm{2D \ square}}(\vec{k}) = -J\cos (k_x) - J\cos(k_y)$, $E_{\mathrm{2D \  tri}}(\vec{k}) = -J\cos (k_x) - J\cos(k_x - k_y\sqrt{3}/2) - J\cos(k_x + k_y\sqrt{3}/2)$, $E_{\mathrm{3D \ square}}(\vec{k}) = -J\cos (k_x) - J\cos(k_y) - J\cos(k_z)$, and $E_{\mathrm{3D \  tri}}(\vec{k}) = -J\cos (k_x) - J\cos(k_x - k_y\sqrt{3}/2) - J\cos(k_x + k_y\sqrt{3}/2) - J\cos(k_z)$ for a 2D square, 2D triangular, a 3D square, and a 3D triangular lattice respectively. The theoretical curves for the density of states were computed using the Kwant code~\cite{Groth2014}. 

In order to take paths through the high symmetry points of the Brillouin zone, high resolution experimental data was needed. For the band structure measurements, the cavity was modulated at a slightly detuned frequency ($\sim 10-20$ Hz) in order to match with a multiple of the sampling rate of the oscilloscope. This also prevented a walk-off in the reconstructions when the scope was sampling rate was detuned from the cavity mode spacing, leading to a linear shift in the band structure along $t_{\mathrm{slow}}$ when digitized.

\subsection{Input-state preparation}
\label{sec:input_state_prep}

\begin{figure}
    \centering
    \includegraphics{ 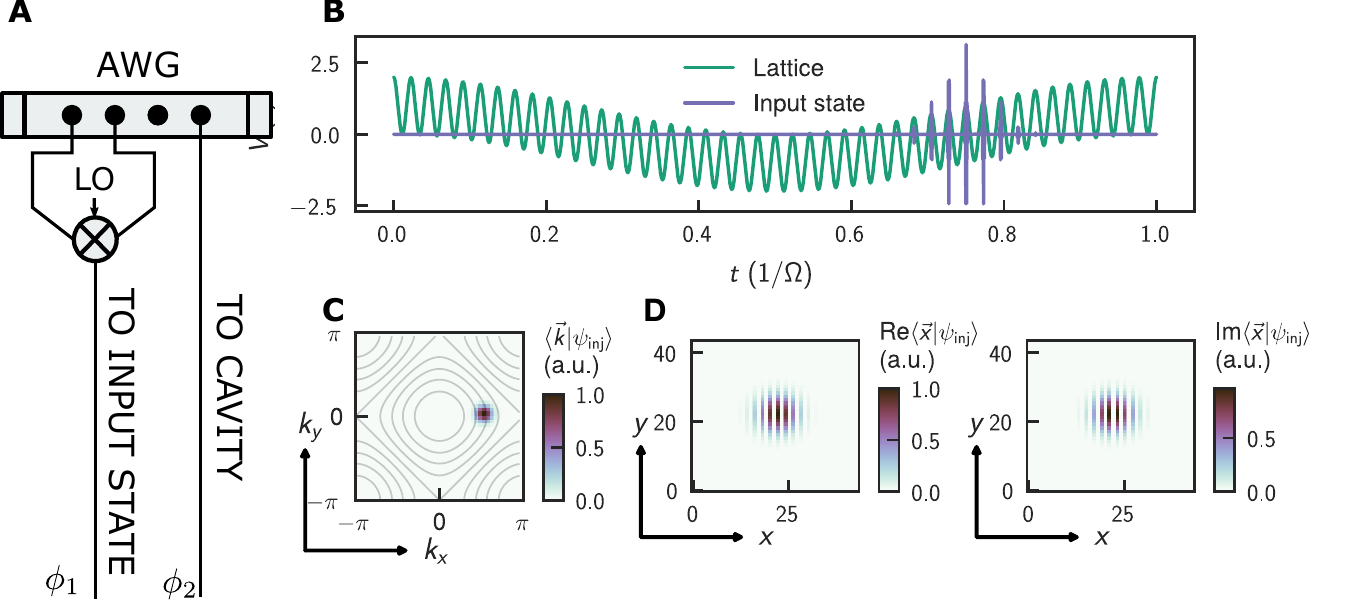}
    
    \caption{Overview of input state preparation in a 2D square lattice. (A) Two channels of an AWG form the in-phase and out of phase quadratures of a signal upconverted with a 12 GHz LO. A third channel is sent to the intra-cavity EOM. These channels are locked to the same clock, therefore the phase difference between the input state preparation and the cavity phase modulation is given by the cable length. (B) Time-domain signal sent to the cavity phase modulator (green) and the injection phase modulator (blue). (C) Two dimensional representation of the input state in momentum/time space with level curves of the 2D band structure  plotted in gray. (D) Real and imaginary parts of the injected signal in coordinate/frequency space.  }
    \label{fig:input_state_prep}
\end{figure}

To prepare arbitrary input states, high modulation bandwidth and high preparation fidelity with no spurious images or modes are required. To this end, we implemented an image rejection IQ mixer in the optical domain by combining a 12 GHz-phase modulator with a fiber Bragg grating as a filter (see Fig.~\ref{fig:exp_setup}). The Bragg grating is a 4 GHz-bandpass filter, which enabled the programmability of around 4000 lattice sites, while rejecting the spurious sidebands in addition to the carrier frequency. The filter is centered 12 GHz away from the injection seed, so the modulator was driven with a signal with a center frequency of 12 GHz. The baseband signal around the 12-GHz sideband was encoded in a voltage signal from an AWG as I and Q pairs. These two were then sent into an (electronic) IQ mixer upconverted with a 12-GHz local oscillator (Fig.~\ref{fig:input_state_prep}). The resultant upconverted signal was then sent to drive the external EOM, enabling both phase and amplitude programmability of the input state at every lattice position. 

\begin{figure}
    \centering
    \includegraphics{ 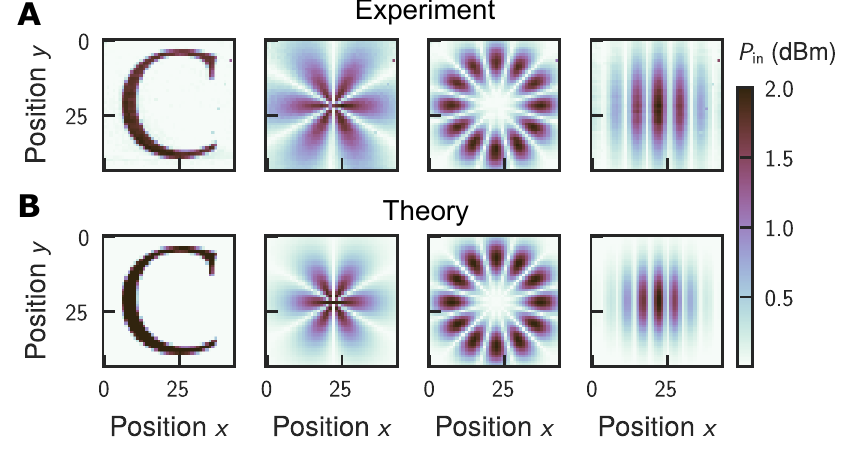}
    
    \caption{Comparison of experimentally prepared input states with the digital image. (A) Experimentally measured input states measured with heterodyne with a detector before the cavity (see Fig.~\ref{fig:figure3} of the main text). Here, a portion of the input spectrum has been folded into a $44 \times 44$ image. (B) Digital image of the input states fed as input into the phase modulator. }
    \label{fig:input_state_prep_data}
\end{figure}

To account for spectral inhomogeneities in the input chain, we calibrated the modulation by measuring the light before entering the cavity. If the modulation has some inhomogeneity, such that a voltage signal $V(t) = \sum_n V_n \sin (n \Omega t)$ is modified to $V(t) = \sum_n (V_n\eta_n) \sin (n \Omega t)$, the spectrum analyzer output modes are scaled by $\eta_n^2$. To compensate, we injected light that had been uniformly modulated at all integer multiples of the FSR within the spectral region of interest, i.e. a top hat distribution defined on latttice sites. This region of interest is roughly up to 2000 modes for the input states presented in Fig.~\ref{fig:figure3}B in the main text. The square root of the response of this measurement gives us approximately $\eta_n$, which we used to apply an envelope function to the modulation signal. If successful, a top hat modulation multiplied with this envelope function will produce a clean flat spectrum. Otherwise, this process can be iterated for higher order. The measured input states shown in Fig.~\ref{fig:input_state_prep_data} were produced with just a single iteration of the above procedure. 

The phase modulator preparing input states and the phase modulator programming the cavity interactions are synced to the same clock, however, the voltage signal driving the input state modulator is upconverted before hitting the EOM (Fig.~\ref{fig:input_state_prep}). This imparts a phase difference between the input state and the cavity due to the different cable lengths. As shown in Fig.~\ref{fig:input_state_prep}B, the phase difference between the injected light and the cavity defines the average momentum of the excitation. To account for this delay, we prepared a state with net-zero momentum (in the frame of the outgoing signal) into a tight-binding lattice, and tuned the phase of the cavity signal until we observed no transport. 

\subsection{Supplementary results}
\label{sec:supp_results}

\begin{figure}
    \centering
    \includegraphics{ 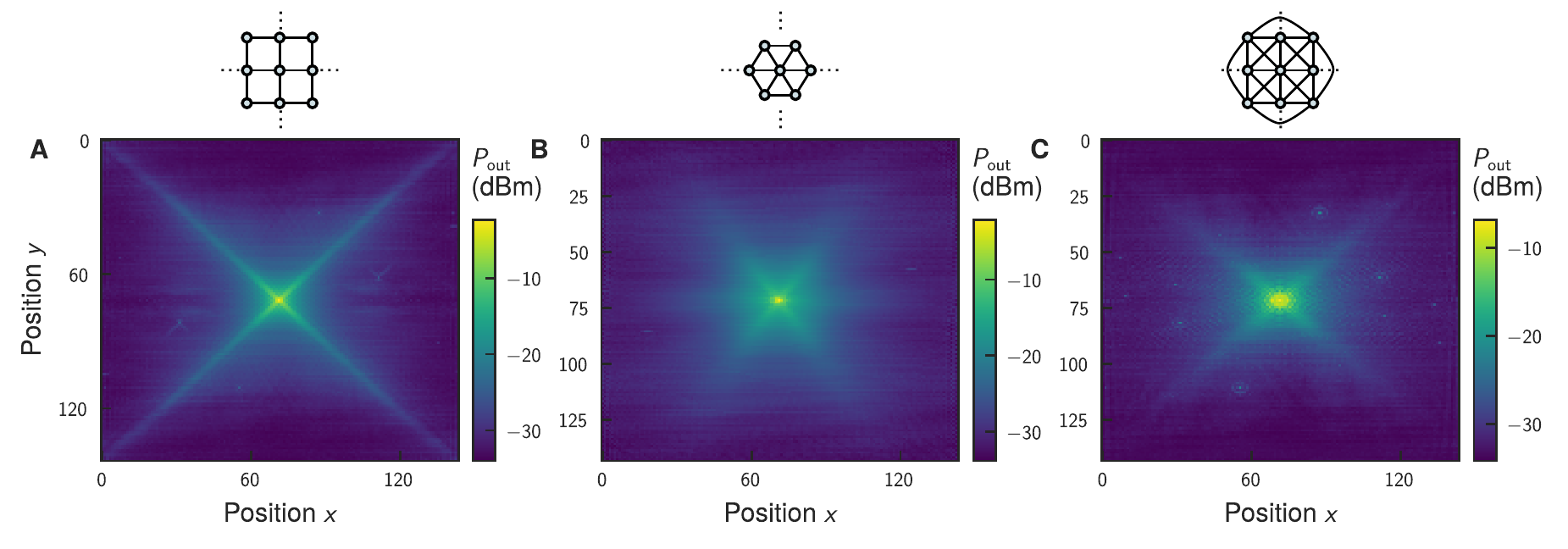}
    
    \caption{Response of single site injection in a 2D square lattice with nearest neighbor connections, 2D triangular, and 2D square with second and third-nearest connections. }
    \label{fig:lattice_ss_response}
\end{figure}

Here we present additional measurements of transport in a variety of lattices, showcasing the programmability of our photonic system. The top row of Fig.~\ref{fig:lattice_ss_response} shows the steady state cavity response for a 2D square lattice with nearest neighbor connections, a 2D triangular lattice, and a 2D square lattice with nearest and next-nearest neighbor connections. Beneath these are the simulations of the corresponding lattices with a tight-binding Hamiltonian without the twisted boundary conditions.

\begin{figure}
    \centering
    \includegraphics{ 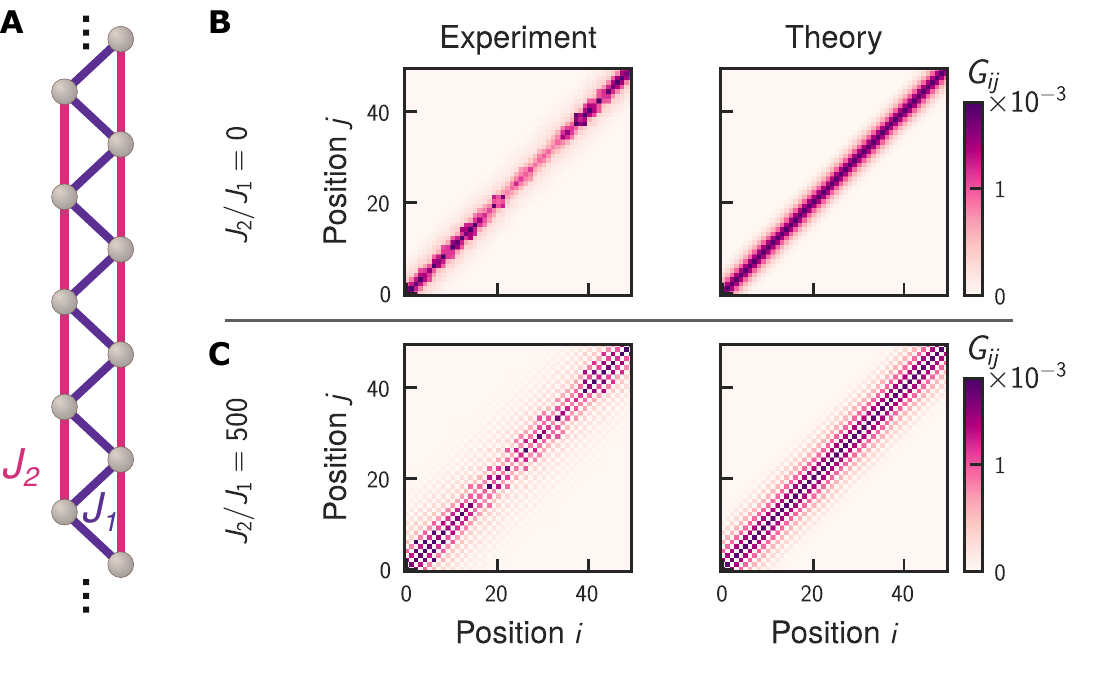}
    
    \caption{A 1D lattice with both nearest neighbor and next-nearest neighbor connections (A). With only nearest neighbor connections (B), the correlations decay exponentially. As the next-nearest hopping is increased (C), the correlation length doubles on average, and nearest neighbor sites become uncorrelated.  }
    \label{fig:j1j2_full}
\end{figure}

Figure~\ref{fig:j1j2_full} shows the correlations of a simple 1D lattice with either nearest neighbor hoppings, or next-nearest neighbor hoppings. We see that in Fig.~\ref{fig:j1j2_full}, as the strength of the nearest-neighbor coupling becomes small relative to the next-nearest neighbor coupling, the correlation length broadens and becomes tessellated, indicating that nearest neighbors are no longer correlated. The inhomogeneity along the diagonals is due components in the readout chain, e.g. detector, amplifier, and spectrum analyzer, as well as input chain, e.g. bandpass filters, amplifier, and injection. These matrices are constructed from the measured spectral response of the lattice at a given injection site. That is,
\begin{equation}
    G_{ij} = \langle n(i)^*n(j) \rangle - \langle n(i)^* \rangle \langle n(j) \rangle,  
\end{equation}
where $n(j)$ denotes the population at site $j$, and $n(i)^*$ denotes the population at site $i$ given the injection was made at site $i$, and the $\langle ... \rangle$ brackets denote the average over injection sites. We denote these quantities as the non-equilibrium correlation matrices, and find these measurements capture correlations of systems also found in other literature \cite{Periwal2021}

\begin{figure}[h]
    \centering
    \includegraphics{ 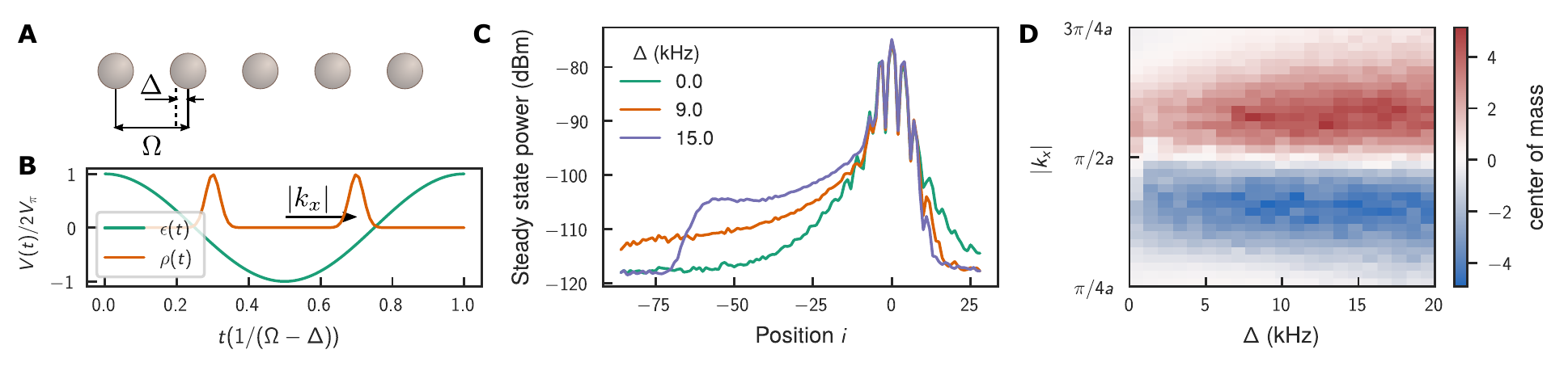}
    
    \caption{ Synthetic electric fields and Bloch oscillations via detuned cavity modulations. (A) Coupling nearest neighbor sites with a detuned drive implements an effective electric field (see text). (B) Green and orange curves are the voltage signals driving cavity and input phase modulators respectively. The input state is chosen to be a superposition of two Gaussian wavepackets with equal and opposite momenta $k_x$, forming a standing wavepacket. (C) Steady state output cavity spectra for different values of detuning. With no detuning (green) the response is symmetric. As detuning is increased, the light will traverse to a maximum amplitude given by the inverse of the detuning. (D) Full experimentally measured phase diagram of Bloch oscillations captured by measuring the center of mass of the output spectra (Eq. \eqref{eq:center_of_mass})}
    \label{fig:bloch_oscillations}
\end{figure}

Finally, we extend previous works in realizing synthetic electric fields via the observation of Bloch oscillations, shown here in a 1D nearest-neighbor chain for arbitrary input states. As in Refs.~\cite{Li2021,Chen2021,Yuan2016}, we realize a synthetic voltage by modulating the cavity with a slightly detuned frequency, as depicted in Fig.~\ref{fig:bloch_oscillations}. A general time-dependent phase modulation for a nearest-neighbor coupled chain realizes the Hamiltonian
\begin{equation}
    H = \sum_n a^{\dagger}_{n+1}a_n e^{i\theta(t)} + \mathrm{H.c.} 
\end{equation}
By performing the gauge transformation $\vert \Psi \rangle = \sum_n  C_n a^{\dagger}_n\vert 0 \rangle \rightarrow \sum_n C_n e^{i\theta(t)}a^{\dagger}_n\vert 0 \rangle$, the Hamiltonian becomes~\cite{Yuan2015}
\begin{equation}
    H = \sum_n (a^{\dagger}_{n+1}a_n + \mathrm{H.c.} ) + \sum_n n \dot{\theta}(t) a_n^{\dagger}a_n.
\end{equation}
By simply detuning the modulation by $\Delta$, such that $\theta(t) = \Delta t$, we implement a linearly increasing voltage, or equivalently, a constant electric field in the 1D chain. These give rise to Bloch oscillations \cite{Yuan2016}, which have been seen in photonic simulators, but are difficult to observe in real material systems, requiring very clean samples. 

As an example, we prepare a 1D lattice and inject superpositions of equal but opposite momentum wavepackets, i.e. standing wavepackets (Fig.~\ref{fig:bloch_oscillations}B). In the absence of an electric field, these states will diffuse, growing the Gaussian envelope of the packet. In the presence of an electric field however, the mirror symmetry of the lattice is broken, and the packet gains some overall momentum, shown in the measurements in Fig.~\ref{fig:bloch_oscillations}C. In Fig.~\ref{fig:bloch_oscillations}D, we measure the center of mass 

\begin{equation}
\langle x \rangle = \int_{\mathrm{lattice}} x  \langle x \vert \psi_{\mathrm{out}} \rangle dx 
\label{eq:center_of_mass}
\end{equation}
over all values of standing wave momenta values, as well as detunings up to 20 kHz. 

The above procedure could be generalized to higher dimensions as well as to time-dependent electric fields. In the results presented so far for two-dimensional lattices, the coupling in the second dimension was just an integer multiple of the first, however, these can be independently detuned from each other, instantiating an electric field in either direction. Additionally, by varying the detuning as a function of time, one can implement AC electric fields by frequency modulating the voltage signal driving the phase modulator over a kHz time scale. 

\end{document}